\documentclass[10pt]{elsarticle}
\usepackage{lineno,hyperref}
\modulolinenumbers[5]
\usepackage{amsthm}
\usepackage{amsmath,bm}
\usepackage{anysize}
\usepackage{amsfonts}
\usepackage{amssymb}
\usepackage{algorithm}
\usepackage{algorithmic}
\usepackage{appendix}
\usepackage{bm}
\usepackage{color,xcolor,colortbl,soul}
\usepackage{multirow}
\usepackage{graphics,graphicx}
\usepackage{caption}
\usepackage{epstopdf} 
\usepackage{eurosym}
\usepackage[ cal = cm,
             scr = euler
           ]{mathalfa}
\usepackage{eufrak}
\usepackage{float}
\usepackage{fancyhdr}
\usepackage{longtable}
\usepackage{mathrsfs}
\usepackage{enumerate}
\usepackage{flushend}
\usepackage[framemethod=tikz]{mdframed}
\usepackage{lipsum}
\usepackage{float}
\usepackage{multicol}
\usepackage{verbatim}
\usepackage{natbib}
\usepackage{caption}
\usepackage{subfig}
\usepackage{geometry}
\usepackage{hyperref}
\usepackage{booktabs}
\usepackage{commath,amsmath}
\usepackage{multicol}
\usepackage{placeins}
\usepackage{siunitx}
\usepackage{gensymb}
\usepackage{tablefootnote}
\usepackage{setspace}

\journal{Mechanics of Materials}

\setcitestyle{authoryear}
\date{}
\usepackage{times}
\begin{document}
\begin{frontmatter}
\title{Representative Volume Element: Existence and Extent in Cracked Heterogeneous Medium}
\author[add1]{Hari Sankar R\fnref{fn1}}
\author[add1]{Harpreet Singh\corref{corrauth}}
\cortext[corrauth]{Corresponding author}
\ead{harpreet@iitgoa.ac.in}
\address[add1]{School of Mechanical Sciences, Indian Institute of Technology, Farmagudi, Goa 401301 India}
\fntext[fn1]{Present address: Tata Advanced Systems Limited, Bengaluru, Karnataka 560105 India }
\begin{abstract} \begin{spacing}{1.35}\small
Acknowledging the ever-increasing demand for composites in the engineering industry, this paper focuses on the failure of composites at the microscale and augmenting the use of multiscale modelling techniques to make them better for various applications. This work aims to increase the representativeness of the volume element by attenuating the mesh and size sensitivities in representative volume element (RVE) modelling. A technique to alleviate mesh sensitivity in RVE modelling is proposed, which equalises the fracture energy observed from computational analysis with the real phenomenon, thereby keeping the response independent of the bandwidth of strain localisation. Based on the hypothesis that ensuring periodicity of strain, in addition to displacement periodicity across the domain boundary and supplementing the capability of periodic boundary conditions (PBCs) to attenuate the size dependency in RVE modelling, a set of modified PBCs (MPBCs) are formulated. One thousand two hundred RVE samples falling into combinations of five fibre volume fractions and four RVE sizes are analysed under transverse loading, and the ability of MPBCs to attenuate the effect of RVE size on the precision of material response, particularly in the inelastic regime, is verified. This work also focuses on various factors affecting damage initiation in 2D composite RVEs. The arrangement of a pair of fibres with their members placed close to each other, such that the angle between the direction of loading and an imaginary line drawn between their centres is less, is observed to make the region between them more favourable to damage. \end{spacing}
\end{abstract}
\end{frontmatter}


\section{Introduction}
\label{Introduction}

Fibre-reinforced composites are significant in weight-sensitive industries such as the aerospace industry and replace conventional materials in various applications. Studying the failure in composites is essential to get more insights into the corrective measures to make them better for multiple applications. In the context of analyses to analyse such phenomena, laws of physics and governing equations are better understood at the fine scales. Still, modelling \color{black}every fine detail \color{black} of the system needs to be improved because modelling an entire system at finer scales is highly expensive in terms of computational power and time. Multiscale modelling is a technique that utilises the information from finer scales to approximate the required constants at a coarser scale without modelling the entire system on finer scales. 
\par Provided that the mechanical properties of the constituent phases of a heterogeneous medium are well characterised, it is possible to bound the mechanical properties of the resulting mixture. The rule of mixtures for stiffness components by \cite{voigt-1889} and the rule of mixtures for compliance components by \cite{reuss-1929} are some initial developments in this direction. Another lead is the elasticity solution of a boundary value problem for an ellipsoidal or spherical inhomogeneity of a material in an infinite matrix of another material \citep{eshelby-1957}. This was further taken up by \cite{hashin_zvi} and \cite{mori_tanaka}. \cite{hashin-shtrikman} proposed the tightest bounds possible for a range of composite moduli for a multi-phase material.
\par In addition to these analytical methods, several numerical methods, also known as computational homogenisation or upscaling methods, have been developed to deduce the coarse-scale equations from well-defined fine-scale equations. The basic concept of homogenisation is to predict the mechanical behaviour of a heterogeneous medium by replacing it with an \textit{equivalent homogeneous continuum}; hence, letting the system be analysed with macro-structural behaviour laws obtained from micro-structural information. Usually, the characteristic dimensions of heterogeneities in composite materials are much smaller than the dimensions of the structural component itself. The periodic repetition of a unit cell or a representative volume element (RVE) can approximate cases with a roughly periodical distribution of heterogeneities in such a scenario. Simply put, the representative volume element (RVE) of a material refers to the smallest part of the material that can represent the material as a whole \citep{kaw-2005}. \cite{li-2019} define the \textit{representativeness} of an RVE as its capability to reproduce the properties obtained from a volume of the material of infinite size in the length scale being considered. 
\par Let a chosen volume element be loaded by prescribing either uniform traction or uniform displacement over its boundary. According to \cite{hill-1963}, the volume element could be considered as an RVE when the difference in predicted effective properties in both cases are sufficiently small. An approach to define RVE, based on the concept of homogenisation and forgoes the statistical variations, has been proposed by \cite{drugan-1996}. \cite{kanit-2003} defined RVE as that volume that is adequately large enough to represent the composite in terms of heterogeneities statistically but small enough to be considered a volume element of continuum mechanics. In addition, they propose that RVEs be such that the spatial averages of stress, strain, or energy fields obtained from RVE must represent the constitutive behaviour of the material at the macroscopic level (see Figure \ref{fig:1}). In this work, we follow this definition of RVE. An RVE, as per this definition, does not attempt to include all the heterogeneities and would be smaller than the one according to the definition in terms of statistical representation. Also, RVEs to analyse nonlinear properties such as fracture energy would be larger than those employed to find linear properties such as apparent stiffness \citep{pelissou-2009}. RVEs being the representatives of the heterogeneous material with enough statistical information of the same, fine-scale problem over the RVE domain is solved to obtain the material's effective properties \citep{singh2017reduced, singh2020strain}. 
\par Moreover, the failure of the material is highly influenced by the microscopic heterogeneities, and further evolution of the cracks depends on the micro-scale details such as the size and distribution of the heterogeneities, matrix and fibre strength, volume fraction etc. \citep{bazant2003scaling, nguyen2010existence, sanchez2013failure}. However, the introduction of the material failure process raises significant theoretical concerns that necessitate careful consideration in developing the RVE framework. Actually, loading/unloading mechanisms and strain localisation processes occurring in various micro-scale locations lead to material degradation seen at the macro-scale. Traditional multi-scale methods are primarily based on volume averaging techniques and assume a consistent distribution of macro-scale strains into the micro-scale domain following accepted homogenisation guidelines for stresses, i.e., a volumetric average stretched over the whole RVE. These suppositions, though, have a very shaky physical significance throughout the failure regime. The homogenised response's lack of objectivity concerning the RVE size when conventional homogenisation techniques are used is another crucial issue that has to be highlighted \citep{blanco2016variational}. \color{black}\cite{bouchedjra2018determination} investigated the elastoplastic macroscopic behaviour for polycrystal materials and proposed criteria for determining the size of RVE subjected to monotonic and cyclic loading conditions. The effect of particle volume fractions, effective properties and morphology of the heterogeneities on the RVE size were studied in detail by \cite{el2021numerical} for the linear structural and thermal constitutive framework. Similarly, \cite{bensaada2022void} performed the RVE-based studies for porous materials and devised the macroscopic effective yield surfaces. 
\color{black} 
\par This study results in techniques to \color{black}improve \color{black} RVE modelling by alleviating the mesh and size sensitivities and is a testament to the role of inhomogeneity arrangement in damage initiation and propagation in 2D composite RVEs. Overall, this paper contains eight sections. Section \ref{RVE to Macroscale Linking} explains the classical homogenisation framework for establishing the link between RVE and macroscale. Section \ref{Constitutive Behavior of Matrix and inhomogeneity} describes the constitutive models used for matrix and fibre domains for simulating initiation and propagation of the failure. Furthermore, section \ref{Extent and Existence of RVE} elaborates on the critical contributors which define the extent and existence of RVE in cracked heterogeneous materials. The RVE studies based on three identified parameters are carried out, which are explained in section \ref{FEM Element Size Dependent Softening Response}, section \ref{RVE Size Dependent Softening Response} and section \ref{Hetrogeneity Distribution Dependent Softening Response}. In the end, section \ref{Conclusions} interprets the conclusions drawn from this work and the future scope of the extension.    
\section{RVE to Macroscale Coupling}
\label{RVE to Macroscale Linking}
\par In physics and mathematics, homogenisation is a technique to study partial differential equations (PDEs), which consist of rapidly oscillating coefficients. 
\begin{figure}[H]
\centering
\includegraphics[width=0.9\textwidth]{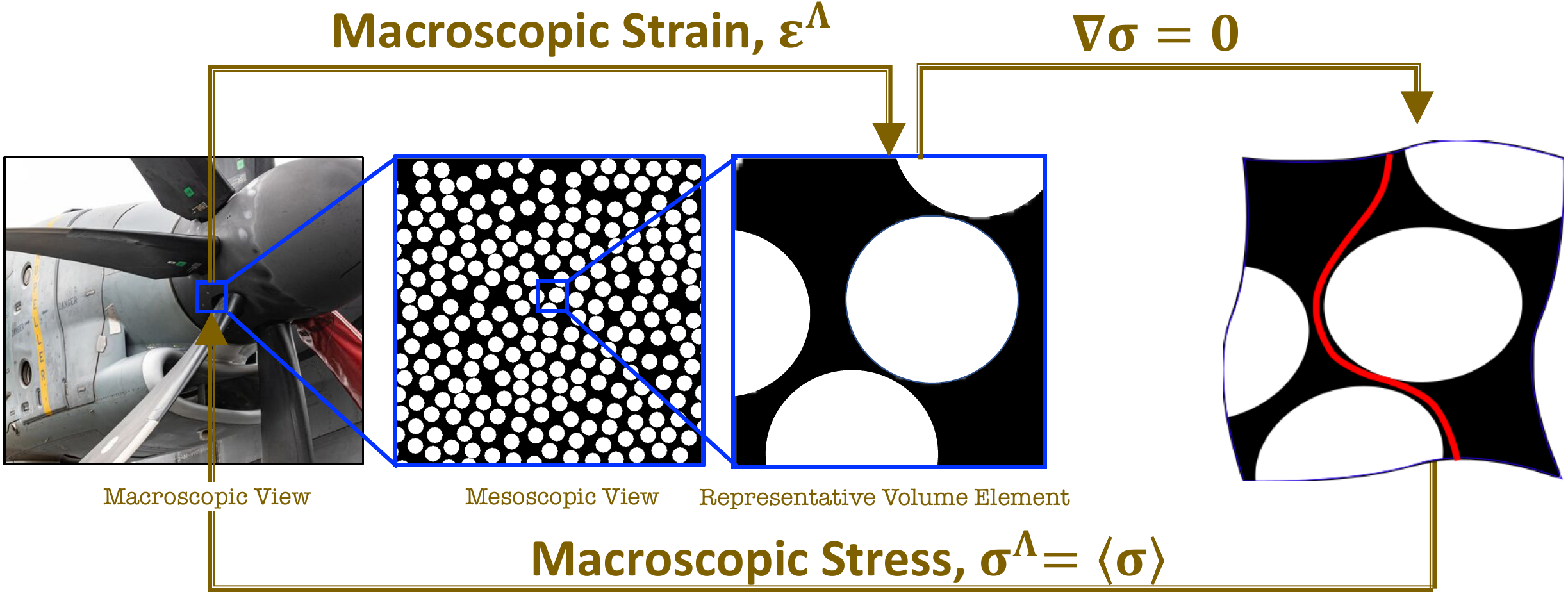}
\caption{\color{black}A typical homogenisation process for a multi-phase material which consists of two scales 1). Macroscale (Scale of laminate/structure) 2). Mesoscale (Scale of heterogeneities). Furthermore, a Representative Volume Element (RVE) is a mesoscopic sub-domain representing the material's constitutive behaviour at the macroscopic level. The information flows from macroscale in terms of macro-strain, $\boldsymbol{\varepsilon}^{\Lambda}$ to RVE, which results in a deformed or cracked configuration and provides homogenised stress, $\boldsymbol{\sigma}^{\Lambda}$ at the macroscale. The red region in the deformed configuration represents the failure.}
\label{fig:1}
\end{figure}
\color{black}
The premise of homogenisation is that the governing equations of individual phases, including their constitutive equations and geometry, are well understood at the fine-scale phases, or at least better understood than at the coarse-scale phases. Under this premise, homogenisation offers a mathematical framework using which coarse-scale equations can be obtained from well-defined fine-scale equations. The advantage of homogenisation is that the behaviour of a heterogeneous material can be determined, at least in theory, without resorting to potentially expensive testing. Figure \ref{fig:1} depicts the coupling of RVE to the macroscale for a multi-phase material.
\par Consider a heterogeneous body without body forces in linear elastic equilibrium. Also, consider a homogeneous body occupying the same space as that of this heterogeneous body and with uniform stress and strain fields. The Hill-Mandel macro-homogeneity condition determines the condition under which the average strain energy density of the heterogeneous body is the same as that of the homogeneous body \citep{hill-1963}. 

\begin{figure}[H]
\centering
\includegraphics[width=0.9\textwidth]{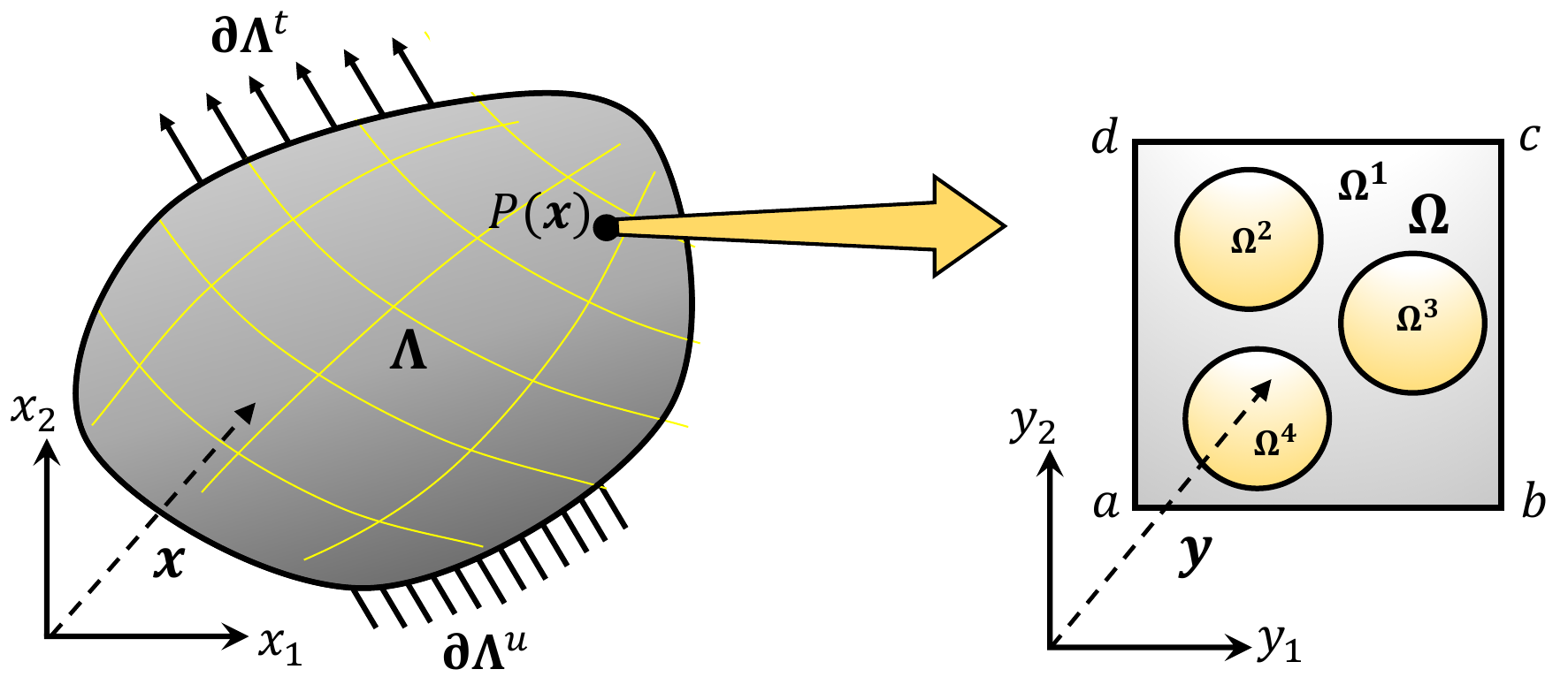}
		\caption[Diagram illustrating scales of damage observation.] {Diagram illustrating coupling between macroscale boundary value problem consisting of macro-domain $\Lambda$ to microscale boundary value problem for micro-domain $\Omega$ (boundary marked as $abcd$). Micro-domain $\Omega$ corresponds to a point denoted as $P(\bold{x})$ of macro-domain $\Lambda$. $\boldsymbol{x}$ represents a position in the macroscale domain, and position representation in the microscale domain is shown as $\boldsymbol{y}$. $\Lambda$ contains the Dirichlet and the Neumann boundaries, shown as $\partial\Lambda^u$ and $\partial\Lambda^t$, respectively. Microscale domain embodies multiple subdomains, shown as $\Omega^1$, $\Omega^2$, $\Omega^3$ and $\Omega^4$ which may be different materials.\label{Fig_3}}
\end{figure}

Let us examine an RVE of volume $\Omega$ and surface $\partial \Omega$, which refers to a material point $P(\boldsymbol{x})$ in the macroscopic domain $\Lambda$ subjected to traction $\boldsymbol{t}$ at boundary $\partial\Lambda^t$ and displacement $\boldsymbol{u}$ at boundary $\partial \Lambda^u$. RVE domain consists of subvolumes $\Omega^1$, $\Omega^2$, $\Omega^3$ etc., as shown in Figure \ref{Fig_3}, and material continuity is maintained between all the subvolumes. The displacement field for this microscale domain is denoted by $\boldsymbol{u}(\boldsymbol{y})$ with components as \{$u_1,$ $u_2,$ $u_3$\} in $\mathbb{R}^3$-Eucleidian space. Similarly, the strain field at the microscale is denoted by a second-order symmetric tensor $\boldsymbol{\varepsilon}(\boldsymbol{y})$, which is often written as $[\varepsilon_{11},$ $\varepsilon_{22},$ $\varepsilon_{33},$ $\gamma_{12},$ $\gamma_{23},$ $\gamma_{13}]^\text{T}$ in the Voigt form. Stress field $\boldsymbol{\sigma}(\boldsymbol{y})$ can be calculated by solving the equilibrium equation for a RVE;
\begin{equation}
	\text{div}\hspace{1mm} \boldsymbol{\sigma} (\boldsymbol{y}) = 0 \label{eq:3}
\end{equation}
when subjected to boundary conditions as per the following
\begin{equation}
	\boldsymbol{u}(\boldsymbol{y})\Big\vert_{\partial \Omega} = \tilde{\boldsymbol{u}}(\boldsymbol{y})\Big\vert_{\partial \Omega} + \bar{\boldsymbol{u}}(\boldsymbol{y})\Big\vert_{\partial \Omega}  \label{eq:4}
\end{equation}

Here the displacement field mentioned earlier is decomposed into two parts. One part $\bar{\boldsymbol{u}}(\boldsymbol{y})$ accounts average displacement field, whereas $\tilde{\boldsymbol{u}}(\boldsymbol{y})$ is for accounting for the fluctuations due to the heterogeneous structure of the domain.
The average displacement field $\bar{\boldsymbol{u}}(\boldsymbol{y})$ is obtained from the macroscale deformation field when subjected to boundary displacements as $\boldsymbol{\varepsilon}^\Lambda\boldsymbol{y}\Big\vert_{\partial \Omega}$, respectively and gives microscopic strain field, $\boldsymbol{\varepsilon}(\boldsymbol{y})$. Similarly, applied surface tractions as $\boldsymbol{\sigma}^\Lambda \boldsymbol{n}\Big\vert_{\partial \Omega}$ gives the macroscopic stress field, $\boldsymbol{\sigma}(\boldsymbol{y})$. Here, $\boldsymbol{x}$ and $\boldsymbol{y}$ correspond to position vectors in the macroscale and microscale domains, respectively. The outward normal to the boundary is denoted by $\boldsymbol{n}\Big\vert_{\partial \Omega}$ and $\boldsymbol{y}\Big\vert_{\partial \Omega}$ is the position vector for the boundary.

Any notation with a bar denotes the volume average of the corresponding quantity. For instance, 
\begin{eqnarray}
	{\boldsymbol{\varepsilon}}^\Lambda &=& \overline{\vphantom{6}\varepsilon}_{ij} = \frac{1}{\vert V_\Omega \vert}\int_{V_\Omega} \varepsilon_{ij}(\boldsymbol{y})dV_\Omega \label{hill_final1}\\
	{\boldsymbol{\sigma}}^\Lambda &=& \overline{\vphantom{6}\sigma}_{ij} = \frac{1}{\vert V_\Omega \vert}\int_{V_\Omega} \sigma_{ij}(\boldsymbol{y})dV_\Omega \label{hill_final2}\\
\overline{\vphantom{6}\sigma_{ij}(\boldsymbol{y})\varepsilon_{ij}(\boldsymbol{y})}-\overline{\vphantom{6}\varepsilon}_{ij}\hspace{1mm}\overline{\vphantom{6}\sigma}_{ij} &=& \frac{1}{|V_\Omega|}\int_{S_\Omega}\left((u_{i}(\boldsymbol{y}) -y_{j}\overline{\vphantom{6}\varepsilon}_{ij})(\sigma_{ik}(\boldsymbol{y})n_{k}(\boldsymbol{y})-\overline{\vphantom{6}\sigma}_{ik}n_{k}(\boldsymbol{y})\right)dS_\Omega
	\label{hill_final}
\end{eqnarray}
where ${\boldsymbol{\varepsilon}}^\Lambda$ and ${\boldsymbol{\sigma}}^\Lambda$ are strain and stress at a point, $P(\boldsymbol{x})$, in the macroscale domain, respectively. The averaging theorems Eq. (\ref{hill_final1}) and (\ref{hill_final2}) establish the coupling between macroscopic and microscopic quantities. Eq. (\ref{hill_final}) is the Hill-Mandel macro-homogeneity condition. Strain energy being half the product of stress and strain, the Hill-Mandel macro-homogeneity condition establishes energy equivalence between the heterogeneous material and the corresponding homogeneous material and is of utmost significance in homogenisation. From Eq. (\ref{hill_final}), it can be inferred that energy equivalence could be achieved between a heterogeneous and a homogeneous body when either of the terms inside the brackets on the RHS becomes zero.
\par PBCs have various advantages over Neumann and Dirichlet boundary conditions. \color{black}Dirichlet BCs overestimate the effective properties, while Neumann BCs underestimate them \color{black}. \cite{ostoja-starzewski-2006} shows that the bounds provided by these two BCs on the effective properties converge as the size of the RVE under consideration increases and that the effective properties predicted with the application of PBCs lie between these two bounds. The values of effective properties predicted from analyses in which PBCs are prescribed lie close to the value of effective property to which the bounds by the other two BCs converge. PBCs give a significant advantage while analysing RVEs of smaller sizes with great accuracy (\citealp{Gitman-2007}; \citealp{canal-2009}; \citealp{totry-2008}; \citealp{totry-2010}; \citealp{ashouri-vajari-2014}). Thus, PBCs are the preferred choice of boundary conditions in RVE modelling. 
\par Periodic boundary conditions (PBCs) in a domain $\Omega$ can be defined on its boundary $\partial\Omega^k$, where subscript $k$ denotes the left $ad$, right $bc$, top $cd$, and bottom $ab$ edges of the domain considered (see Figure \ref{Fig_3}), as:
\begin{eqnarray*}
	\tilde{\boldsymbol{u}}(\boldsymbol{y})\bigg\vert_{\boldsymbol{y}\in\partial\Omega^{ab}} &=& \tilde{\boldsymbol{u}}(\boldsymbol{y})\bigg\vert_{\boldsymbol{y}\in\partial\Omega^{cd}}\\
	\tilde{\boldsymbol{u}}(\boldsymbol{y})\bigg\vert_{\boldsymbol{y}\in\partial\Omega^{bc}} &=& \tilde{\boldsymbol{u}}(\boldsymbol{y})\bigg\vert_{\boldsymbol{y}\in\partial\Omega^{ad}}
\end{eqnarray*}
\section{Constitutive Behavior of Matrix and Inhomogeneity}
\label{Constitutive Behavior of Matrix and inhomogeneity}
\subsection{Continuum Damage Mechanics (CDM) based Framework }
\label{sn: Damage Variable}
\par Consider a surface element at an arbitrary point $P(\boldsymbol{x})$ in an RVE. Once damage occurs, it leaves traces on the volume element in the forms of cracks and cavities, leading to a reduction of the effective area of resistance, say, from $S$ to $\tilde{S}$ \citep{lemaitre1994mechanics} as
\begin{equation}
	\tilde{S} = S -dS
	\label{reduction_S}
\end{equation}
where $dS$ is the void area. This area, $\tilde{S}$, may be considered as the \textit{effective area}, which carries the internal force on the surface element $S$ \citep{kachanov-1986}.
Thus, a damage variable, $D$, may be expressed as 
\begin{equation}
	D(\boldsymbol{x},\:\boldsymbol{n}) = 1 - \frac{\tilde{S}}{S} = \frac{dS}{S}
	\label{D_main}
\end{equation}
where $D(\boldsymbol{x},\:\boldsymbol{n})$ is interpreted as the surface density of discontinuities in the matter in the plane normal to $\boldsymbol{n}$, i.e., it is the decrease in the effective area due to damage development. $D = 0$ and  $D = 1$ denote the initial undamaged and the final completely damaged states, respectively. The direction dependency of the damage variable does not exist for isotropic materials, and the scalar $D$ completely characterises the damaged state. $D$ for isotropic materials is not a function of the orientation $\boldsymbol{n}$. Characterising damage as a function of the surface density of discontinuities leads to the concept of \textit{effective stress}, which is the stress calculated over the area that effectively resists force. As per the strain equivalence principle, total strain ($\boldsymbol{\varepsilon}$) in damaged and effective configurations are equal \citep{murakami-2012}; 
\begin{equation}
	\boldsymbol{\tilde{\varepsilon}}=\boldsymbol{\varepsilon} \label{eq:5}
\end{equation}
Superscript ($\boldsymbol{\tilde{\bullet}}$) represents the effective configuration. 
The nominal and effective stress is given as 
\begin{eqnarray}
	\boldsymbol{\sigma}=\mathcal{L}:{\boldsymbol{\varepsilon}}    \qquad    & &\textsf{Undamaged State}\\
	\boldsymbol{\sigma}=\mathcal{L}(\mathcal{D}):{\boldsymbol{\varepsilon}}               \qquad    & &\textsf{Damaged State} \label{constitutive}
\end{eqnarray}
where $\mathcal{L}$ is a fourth-order isotropic elasticity tensor and $\mathcal{D}$ is a fourth-order damage tensor. Symbol (:) is the double contraction operator. $\mathcal{L}$ can also be represented as 
\begin{equation}
	\mathcal{L}=2G\delta_{ik}\delta_{jl}+\left(K-\dfrac{1}{3}\right)\delta_{ij}\delta_{kl} \label{eq:8}
\end{equation} 
$G$ and $K$ are shear and bulk modulus, respectively and $\delta_{ij}=1 $ if $i=j$ otherwise $\delta_{ij}=0 $. Nominal and effective stress states are expressed as 
\begin{equation}
	\boldsymbol{\tilde{\sigma}}=\left[\mathcal{L}:(\mathcal{L}(\mathcal{D}))^{-1}\right]:\boldsymbol{\sigma} \label{eq:9}
\end{equation}  
Transformation of undamaged constitutive tensor, $\mathcal{L}$ to damaged constitutive tensor $\mathcal{L}(\mathcal{D})$ is given as 
\begin{equation}
	\mathcal{L}(\mathcal{D})=(\mathbb{I}-\mathcal{D}):\mathcal{L} \label{eq:10}
\end{equation}
Here, $\mathbb{I}$ is the fourth-order unit tensor. Using Eq. (\ref{eq:10}), Eq. (\ref{eq:9}) is expressed as 
\begin{equation}
	\boldsymbol{\tilde{\sigma}}=\mathcal{M}:\boldsymbol{\sigma} \label{eq:11}
\end{equation}
where $\mathcal{M}=(\mathbb{I}-\mathcal{D})^{-1}$ is damage effect tensor. Considering the isotropic damage state in the matrix and inhomogeneity $\mathcal{M}$ can be denoted in terms of a single damage variable, $D$;
\begin{equation}
	\mathcal{M}=(1-D)^{-1}\mathbb{I} \label{eq:12}
\end{equation}
By representing fourth order tensor $\mathcal{M}$ as symmetric tensor and in Voigt representation as $6\times 6$ matrix $[M]$;
\begin{equation}
	[M]=\begin{bmatrix}
\frac{1}{1-D} & 0 & 0 & 0 & 0 & 0\\
0 & \frac{1}{1-D} & 0 & 0 & 0 & 0\\
0 & 0 & \frac{1}{1-D} & 0 & 0 & 0\\
0 & 0 & 0 & \frac{1}{1-D} & 0 & 0\\
0 & 0 & 0 & 0 & \frac{1}{1-D} & 0\\
0 & 0 & 0 & 0 & 0 & \frac{1}{1-D}
\end{bmatrix} \label{eq:13}
\end{equation}
\subsection{Damage Threshold and Activation Function}
\label{sn: Damage Threshold and Activation Function}
A damage threshold is defined as that value for the thermodynamic force below which damage does not occur \citep{barbero-2007}. This damage threshold marks off the elastic domain. Damage does not grow when the load state is within this domain. Once the load state gets higher than the threshold, damage increases, the threshold changes, and the evolution of the elastic domain into hardening or softening may occur. A dissipation potential function, $\mathfrak{F}^D$, accounts for the growth of damage.
\par The \textit{damage dissipation function} $\mathfrak{F}^D(Y,\hspace{1mm} D)$ defines as
\begin{equation}
	\mathfrak{F}^D(Y,\hspace{1mm} D) = \dfrac{Y}{(1-D)}\left(\dfrac{\kappa_F}{\kappa_F-\kappa_D}\right)\dfrac{1}{\kappa_D} 
	\label{damageactivation}
\end{equation}
where $Y$ is an independent variable. $\kappa_D$ is the threshold value of strain for initiation of damage, and $\kappa_F$ corresponds to strain for the state of complete failure/damage. $\kappa_D$ is a positive value function dependent on the independent variable and the updated damage threshold for isotropic damage, respectively. Following the positive dissipation principle, which dictates that an irreversible process such as crack formation will always raise the entropy of the system, the updated damage threshold $\kappa_D$ can be written as the sum of the virgin damage threshold $\kappa_{D0}$ and the damage increment $\dot{D}$ which in turn is a function of the damage multiplier function $\dot{\zeta}^D$. Further, the damage evolution can be written as 
\begin{equation}
	\dot{D}=\dot{\zeta}^D\dfrac{\partial \mathfrak{F}^D}{\partial Y}	\label{eq:36}
\end{equation}
where $\dot{\zeta}^D$ is expressed as  
\begin{equation}
	\dot{\zeta}^D=(1-D)\dot{\kappa} \boldsymbol{\mathcal{H}} (\kappa-\kappa_D)	 \label{eq:37}
\end{equation}
$\boldsymbol{\mathcal{H}}$ is Heaviside step function and $\kappa$ is maximum value of the principal strain ($\text{max}\{\langle\hat{\epsilon}_1\rangle, \langle\hat{\epsilon}_2\rangle, \langle\hat{\epsilon}_3\rangle\}$). 
\par
\begin{figure}[H]
\centering
\includegraphics[width=0.7\textwidth]{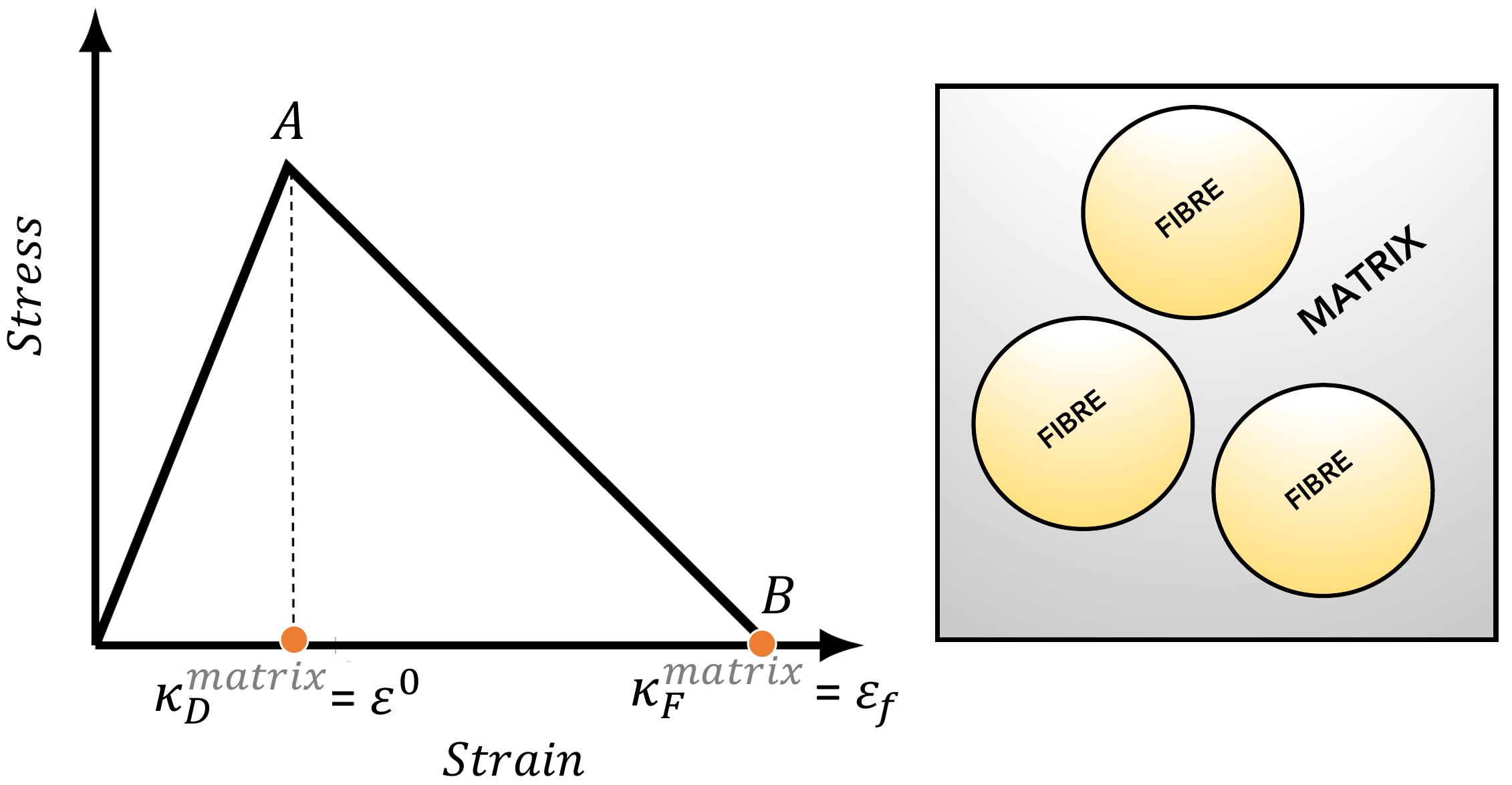}
	\caption[Sample stress-strain diagrams of a quasi-brittle material and a brittle material.]{Typical stress-strain diagram of a strain-softening material for matrix domain. Here,  point $A$ refers to the damage initiation state or ultimate strength and point $B$ denotes the complete failure state. $\kappa_D$ is the damage initiation parameter, equal to damage initiation strain $\varepsilon^0$, and $\kappa_F$ is the failure parameter which is the same as the failure strain $\varepsilon_f$. Overall material exhibits the bi-linear stress-strain variation from start to $A$ and from $A$ to $B$.  \label{quasibrittle_samplecurve}}
\end{figure}

 The Kuhn-Tucker conditions \citep{simo-2006} are defined in terms of the damage multiplier $\dot{\zeta}^D$ and the damage activation function $\mathfrak{F}^D$ as
\begin{equation}
	\dot{\zeta}^D\geq0,\qquad \mathfrak{F}^D\leq0, \qquad \dot{\zeta}^D\mathfrak{F}^D=0 
	\label{kuhn_tucker}
\end{equation}
We extend this framework to identify damage initiation in composite RVE analyses considered in our study. Damage is detected to be initiated in a constitutive phase of the composite when the principal strain at a point in that phase becomes greater than or equal to the damage initiation strain of that phase, as shown in Figure \ref{quasibrittle_samplecurve}. This damage initiation law is incorporated into the analysis using a user-defined subroutine characterising the material behaviour. This work uses the damage initiation criterion mentioned in this section to detect damage initiation and the linear damage evolution law Eq. (\ref{eq:37}) to consider the growth of the damage variable.
\par The damage initiation parameter $\kappa_D$ is known a priori and fed as the damage initiation criterion input. Once the principal strain becomes greater than or equal to the damage initiation strain of any constitutive phase, damage initiates in that phase. In this sample stress-strain diagram, point $A$ corresponds to the damage initiation state of the material. Until point $A$, the damage state variable $D$ is zero. Once the damage initiates, the material's stiffness degrades from points $A$ to $B$.
\par Further, the magnitude of the damage variable increases as per Eq. (\ref{eq:36}). The reduction in material stiffness with increasing strain, as observed from points $A$ to $B$, is known as strain-softening and is a consequence of heterogeneities and brittleness of the material \citep{bazant-1984}. Finally, the material fails at the critical or failure strain $\kappa_F$. At this fully damaged state denoted by point $B$ in \figurename~\ref{quasibrittle_samplecurve}, $D = 1$ corresponds to a state of complete loss of stiffness.
\section{Extent and Existence of RVE}
\label{Extent and Existence of RVE}
\begin{figure}[H]
\centering
\includegraphics[width=0.9\textwidth]{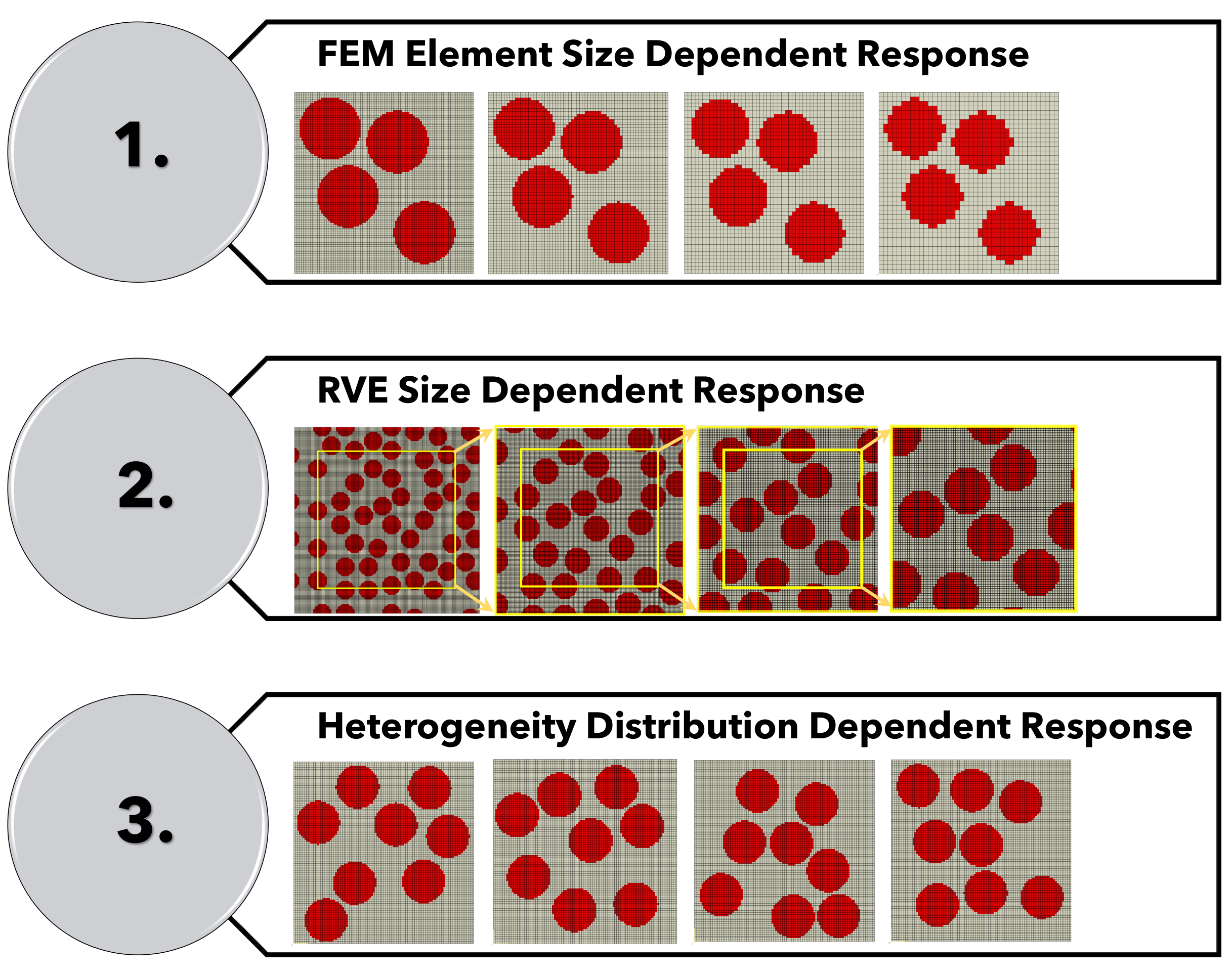}
	\caption[Sample stress-strain diagrams of a quasi-brittle material and a brittle material.]{This figure illustrates the three important parameters that strongly influence the macroscale stress-strain behaviour obtained from averaging the field variables of representative volume element: 1). FEM element size dependent softening response (as described in Section \ref{FEM Element Size Dependent Softening Response}) 2). RVE size-dependent softening response (as described in Section \ref{RVE Size Dependent Softening Response}) and 3). Heterogeneity distribution dependent softening response (as described in Section \ref{Hetrogeneity Distribution Dependent Softening Response}) \label{parameter}}.
\end{figure}

The representativeness of an RVE lies in its capability to provide better approximations of the macroscale behaviour. But, microscale modelling is challenged by various sensitivities, \color{black} as shown in Figure \ref{parameter}\color{black}, that can introduce further variations in the mechanical response of the RVE from that of the corresponding macroscopic component. The results of FE analysis are highly dependent on the mesh size used, and mesh sensitivity is a prolonged issue in RVE modelling. One of the objectives of this work is to assimilate the root cause of this mesh dependency concerning microscale damage analysis and to propose a technique to alleviate mesh sensitivity in RVE modelling. \color{black} The proposed methodology \color{black} will assist in performing analysis at different mesh sizes and obtaining accurate estimates of the material response irrespective of the mesh size used. Also, it allows us to perform analysis with coarser mesh sizes and still obtain results from a study conducted using a finer mesh.
\par Since the purpose of microscale damage analysis concerning multiscale modelling of failure in composites is to get better estimates of properties to be substituted in the coarse-scale problem, RVEs in such a context mean the smallest volume element that could accurately reproduce the macroscopic properties. Describing an RVE as that volume adequately large enough to statistically represent the composite in terms of heterogeneities\footnote{Defining an RVE as that volume, which when repeated in all the directions in space results in the macroscopic component.} would drastically increase its size and hence, the computational expenses to analyse them, and is unnecessary in such a context. There is enough evidence from past investigations that the accuracy of predicted estimated properties diminishes with a decrease in RVE size. But, researchers have also demonstrated the capability of prescription of BCs such as PBCs \citep{ostoja-starzewski-2006} and imposition of geometrical constraints such as the absence of \textit{wall-effects} \citep{Gitman-2007} to increase the accuracy of predicted properties from RVEs of smaller size. We focus on studying the size sensitivity in RVE modelling and propose a new set of PBCs to attenuate mesh sensitivity.
\par The material response of composite microstructures having the same composition and mechanical properties would be almost identical in the elastic region. But, the material behaviour from damage initiation varies noticeably due to the localisation in terms of damage, which occurs due to the position of heterogeneities in the microstructure \citep{pelissou-2009}. Clarity on various factors in fibre arrangement that makes a location favourable to damage will help to predict the damage initiation location in an RVE by looking into its fibre distribution. Depending on the greatest extent to which any of the regions in each RVE favours damage, predictions could be made on the order in which a given set of RVEs experience damage initiation. The favourability of regions within the RVE to damage could be reduced by utilising the knowledge of factors affecting damage initiation and strengthening the RVE with the slightest modifications in fibre arrangement. As the carpenter's rule goes, ``measure twice, cut once''. Modelling what to analyse is as essential as how to study. Insights into various factors that determine the material behaviour post-damage initiation would contribute towards better algorithms to model composite microstructures that can withstand a higher load without experiencing damage initiation. So, our research also focuses on determining the factors affecting damage initiation in 2D composite RVEs.
\section{FEM Element Size Dependent Softening Response}
\label{FEM Element Size Dependent Softening Response}
\begin{figure}[H]
	\centering
\includegraphics[width=0.5\textwidth]{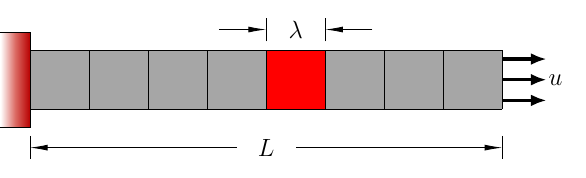}
	\caption[A 1D rod of length $L$ considered to explain strain localization]{A 1D rod of length $L$ considered to explain strain localization. After damage initiation, the stress is assumed to decline in damage unloading in a domain of length $\lambda$ and in elastic unloading in a domain of length $L - \lambda$.}
	\label{1Drod_strain_localization}
\end{figure}

Strain softening induces localisation of mechanical fields such as strain and damage into narrow regions within the material domain as experimentally verified by \citep{desrues}. Consider a 1D brittle rod fixed at one end and subjected to an increasing displacement $u$ on the other end as shown in \figurename~\ref{1Drod_strain_localization}. Once strain crosses its threshold value $\varepsilon^{0}$, the stress does not increase anymore and is assumed to decline in two branches; damage unloading in a domain of length $\lambda$ and elastic unloading in a domain of length $(L-\lambda)$. Then, the stress can be determined as
\begin{equation}
	\sigma = E\varepsilon^{0} \left( \dfrac{\frac{u}{L\varepsilon^{0}} - \frac{\lambda}{L}\left(1+\frac{E}{H}\right)}{1-\frac{\lambda}{L}\left(1+\frac{E}{H}\right)}\right)
	\label{stressdependencylambda}
\end{equation}
where $E$ denotes the material's elastic modulus, and $H$ is the slope of the stress-strain curve in the softening region. Eq. \eqref{stressdependencylambda} depicts the dependency of material response on the length of the damaged zone.
When the domain is discretised into several finite elements for FE analysis, the length of the damaged zone $\lambda$ will necessarily be a multiple of the size of a single element since the strain is approximated as piecewise continuous functions as soon as any element hits the threshold before the other ones, leading to localisation of damage in a narrow region of width precisely equal to that of a single element. The bandwidth of strain localisation $\lambda$ is a material property shown by \cite{bazant-1989}, but it varies depending on the mesh size while performing FE analyses.

Strain localisation leads to variation in the damaged area, resulting in energy dissipated during damage among the analysis results. Since the observed value of energy dissipated during damage varies depending on the mesh, the $F$-$d$ curves of the same sample, when analysed at different mesh sizes, also vary. This section of the paper proposes a technique to alleviate mesh sensitivity in RVE modelling by equalising the observed energy dissipation during FE analysis to that in the actual phenomenon. 
\par
\begin{figure}[H]
	\centering
	\subfloat[\label{undeformed_XX_FS_1_5_a}]{\includegraphics[width=0.45\textwidth]{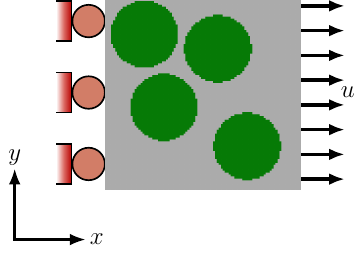}}\hspace{0.1cm}\\	
	\subfloat[Element size 0.010\label{undeformed_XX_FS_1_5_e}]{\includegraphics[width=3.25cm]{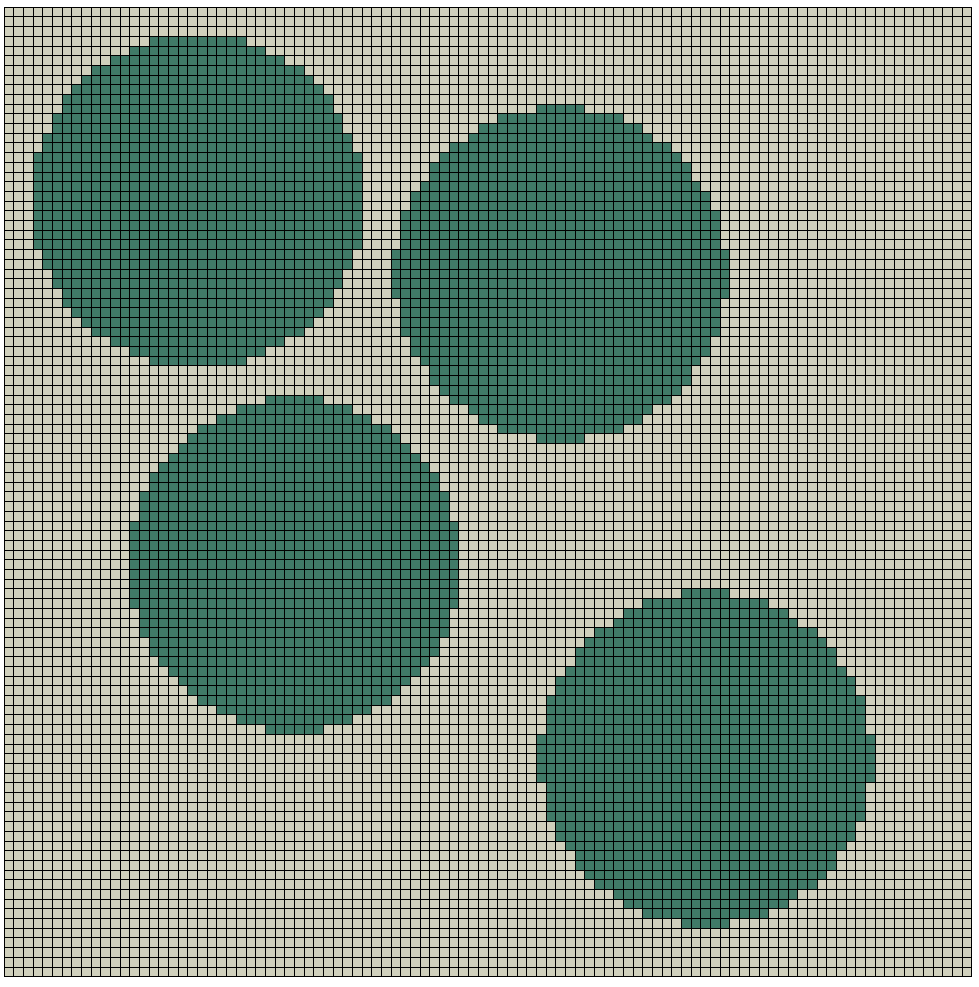}}\hspace{0.09cm}
	\subfloat[Element size 0.015\label{undeformed_XX_FS_1_5_f}]{\includegraphics[width=3.25cm]{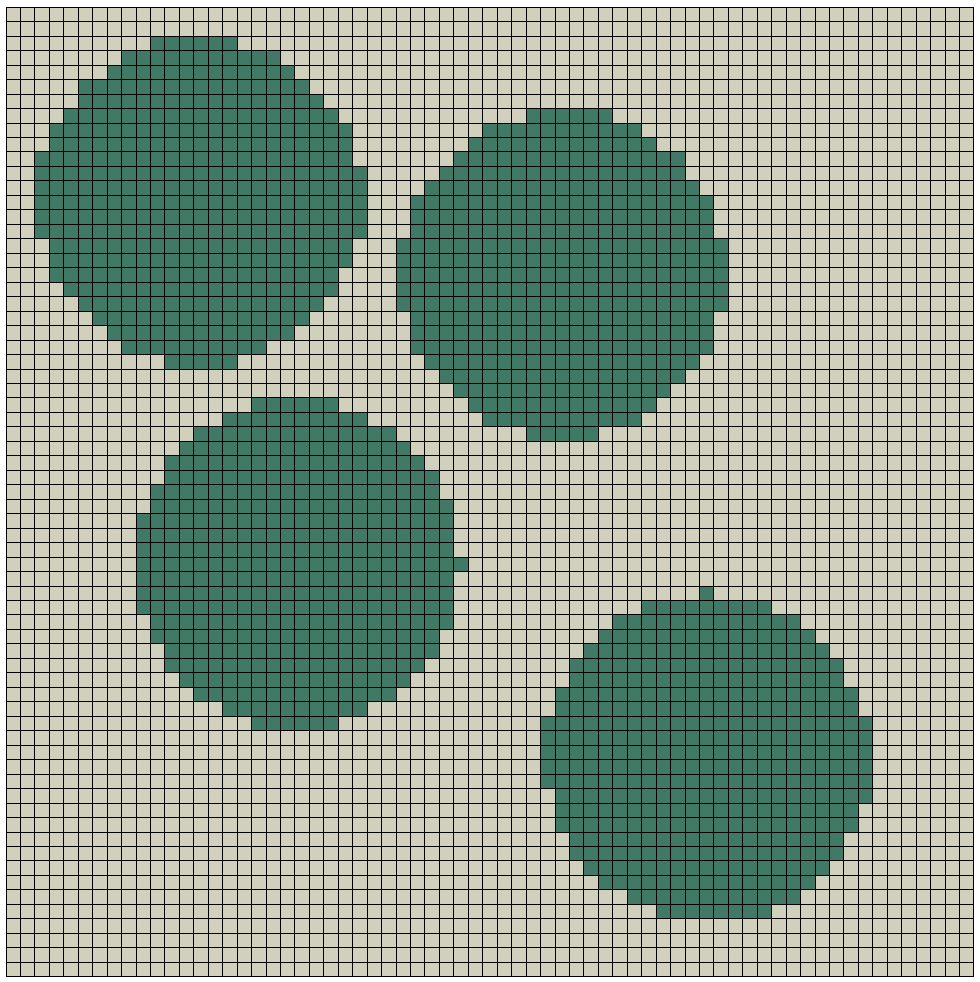}}\hspace{0.09cm}
	\subfloat[Element size 0.020\label{undeformed_XX_FS_1_5_g}]{\includegraphics[width=3.25cm]{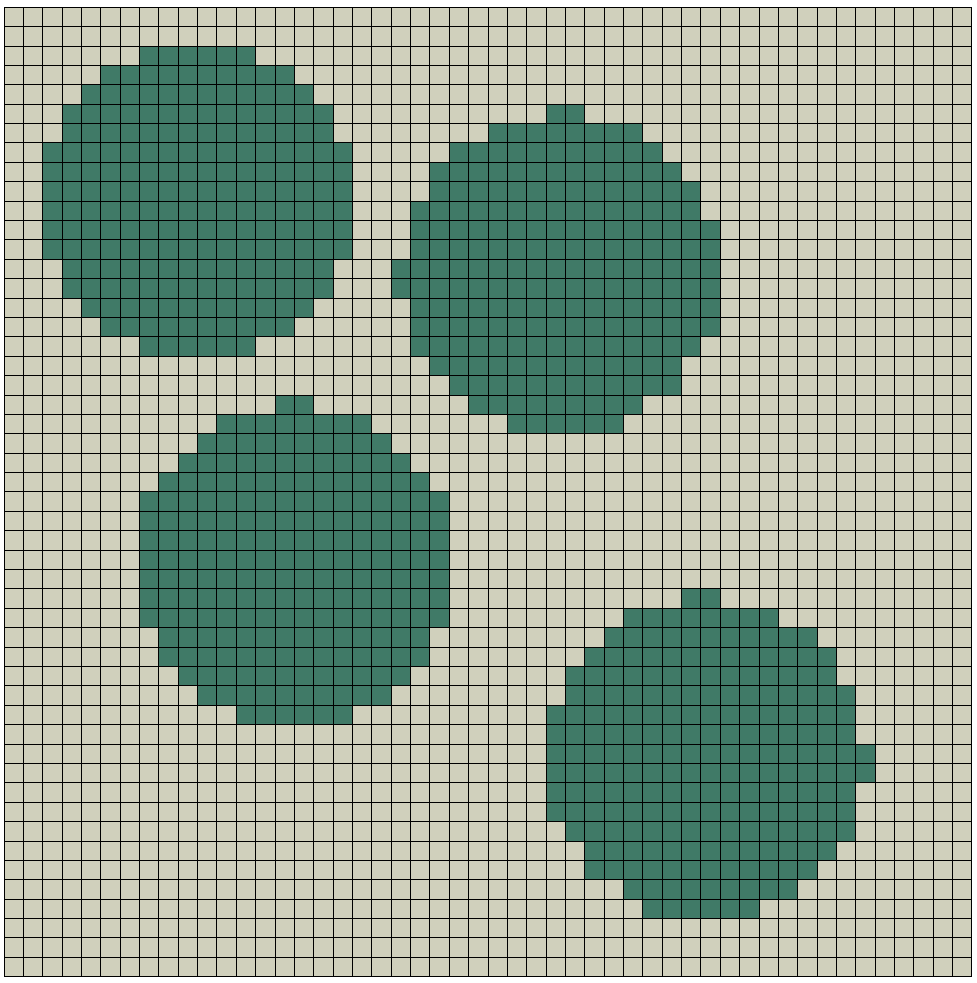}}\hspace{0.09cm}
	\subfloat[Element size 0.025\label{undeformed_XX_FS_1_5_h}]{\includegraphics[width=3.25cm]{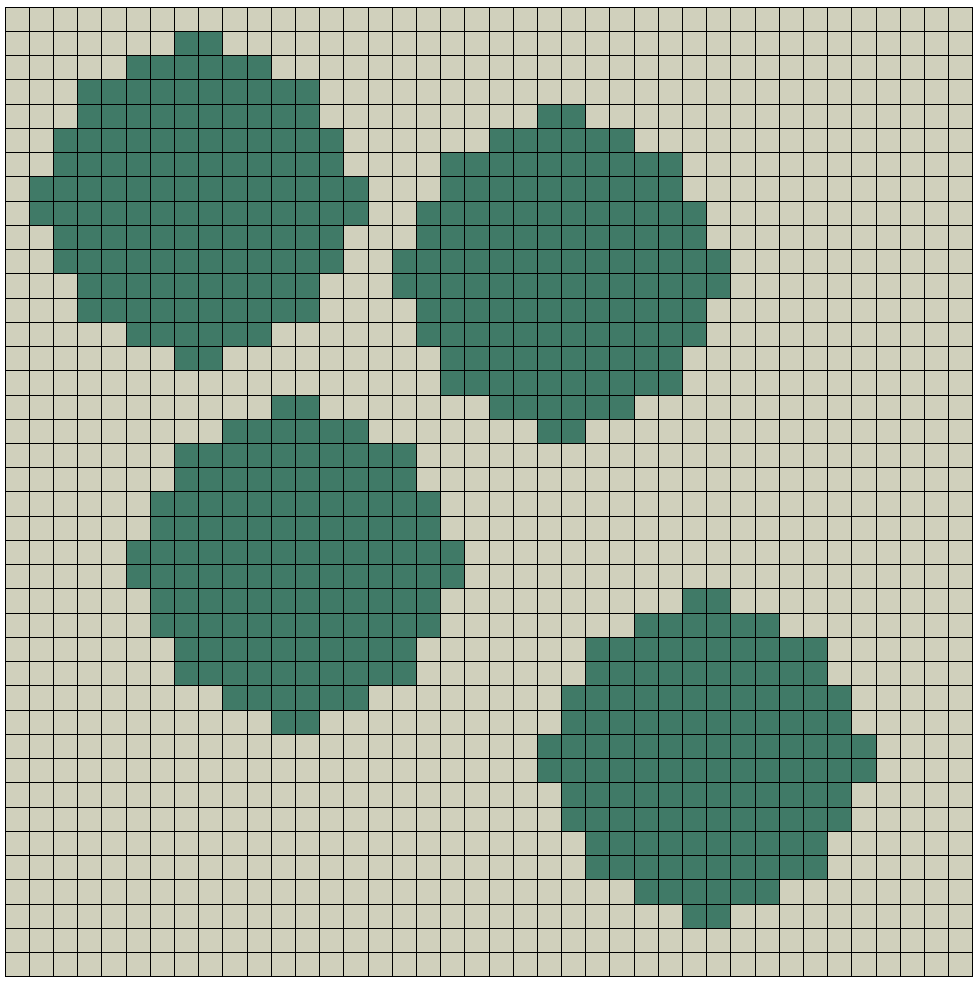}}\\
	\subfloat[Element size 0.010\label{undeformed_XX_FS_1_5_a}]{\includegraphics[width=3.25cm]{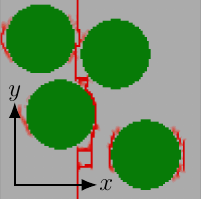}}\hspace{0.1cm}
	\subfloat[Element size 0.015\label{undeformed_XX_FS_1_5_b}]{\includegraphics[width=3.25cm]{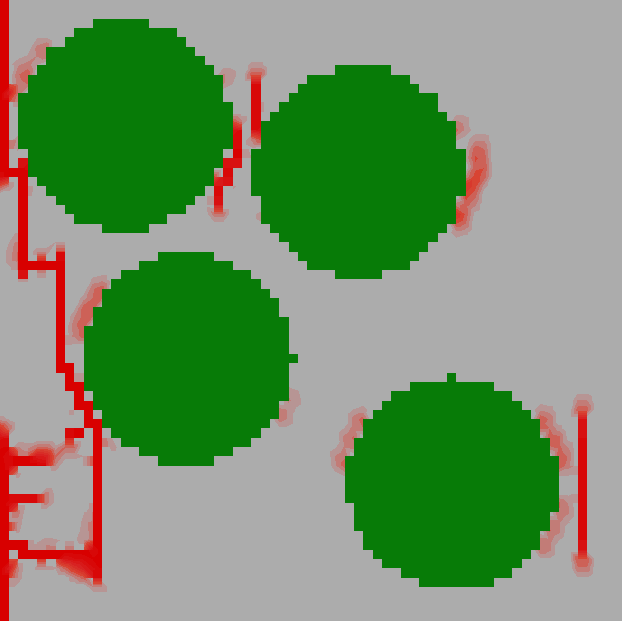}}\hspace{0.1cm}
	\subfloat[Element size 0.020\label{undeformed_XX_FS_1_5_c}]{\includegraphics[width=3.25cm]{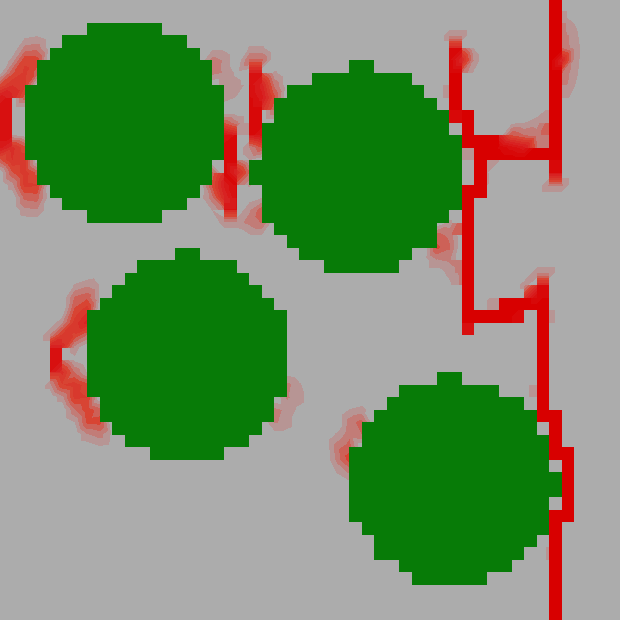}}\hspace{0.1cm}
	\subfloat[Element size 0.025\label{undeformed_XX_FS_1_5_d}]{\includegraphics[width=3.25cm]{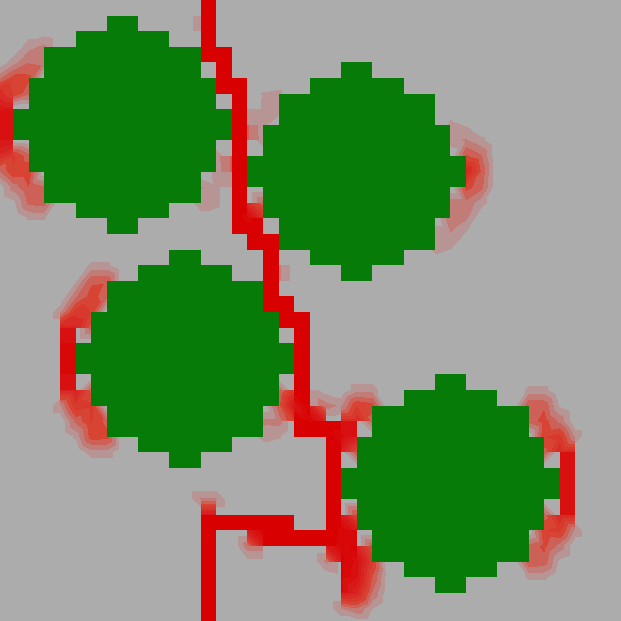}}
	\caption{(a) RVE considered studying the effects of mesh dependency on material response. (b), (c), (d) and (e) illustrate the meshed domain with element sizes 0.01, 0.015, 0.02 and 0.025, respectively. A uniform element size is used for each case. (f), (g), (h) and (i) depict the crack paths (shown on the undeformed shape of the RVE) corresponding to $xx$ loading case (as shown in (a)), with failure strain of the material assigned as 1.5 and with element sizes 0.01, 0.015, 0.02 and 0.025, respectively.}
	\label{undeformed_XX_FS_1_5}
\end{figure}

 To check the mesh sensitivity in RVE modelling, we perform analyses on a 2D RVE of unit size consisting of four fibres and a total fibre volume fraction of 0.37. The fibres are defined by selecting elements and assigning them the fibre properties rather than defining geometric boundaries. The load is applied in the form of displacement as shown in Figure \ref{undeformed_XX_FS_1_5}(a).  
\par On the constrained edge and the edge on which load is applied, displacement periodic boundary conditions are applied in the direction other than that in which load is applied. On the other two edges, PBCs are applied in both directions. 
\par The damage initiation criteria described in section \ref{sn: Damage Threshold and Activation Function} is being utilised to evaluate damage initiation, and linear damage evaluation law is being followed. The sum of the reaction forces at the nodes on the fixed edge and the displacement (in the direction in which the load is applied) of any nodes on the opposite edge is extracted. Post-processing, this data is used to plot the $F$-$d$ diagram. $F$-$d$ plot represents the macroscopic $\sigma$-$\epsilon$ variation because of the unit dimensions of the RVE.  
\par
\begin{figure}[H]
	\centering
	\subfloat[Failure strain = 1.5\label{XX_FS_1a}]{\includegraphics[trim = 0 0 0 0, clip, height=3.5cm]{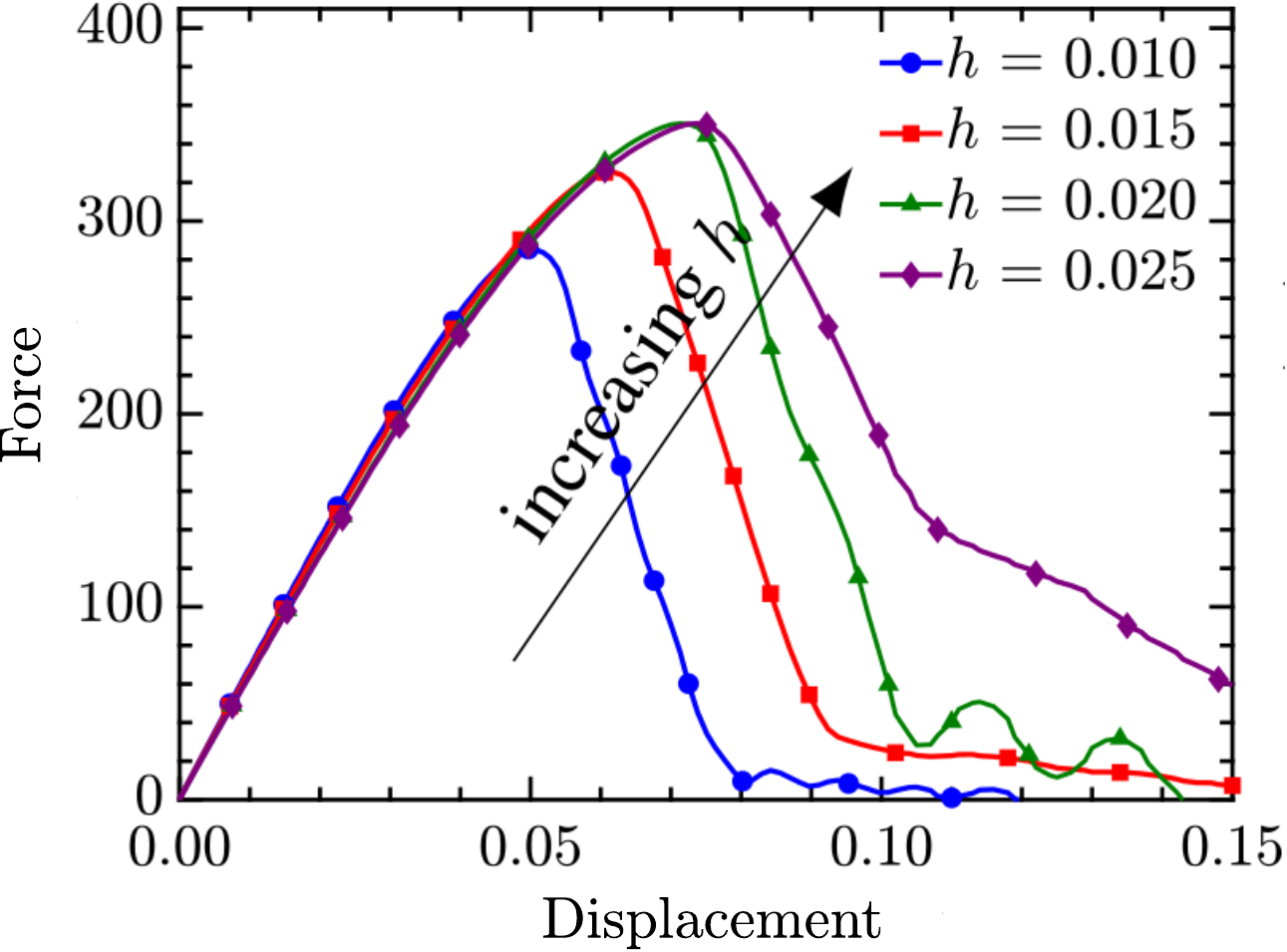}}
	\subfloat[Failure strain = 2.0\label{XX_FS_1b}]{\includegraphics[trim = 0 0 0 0, clip, height=3.5cm]{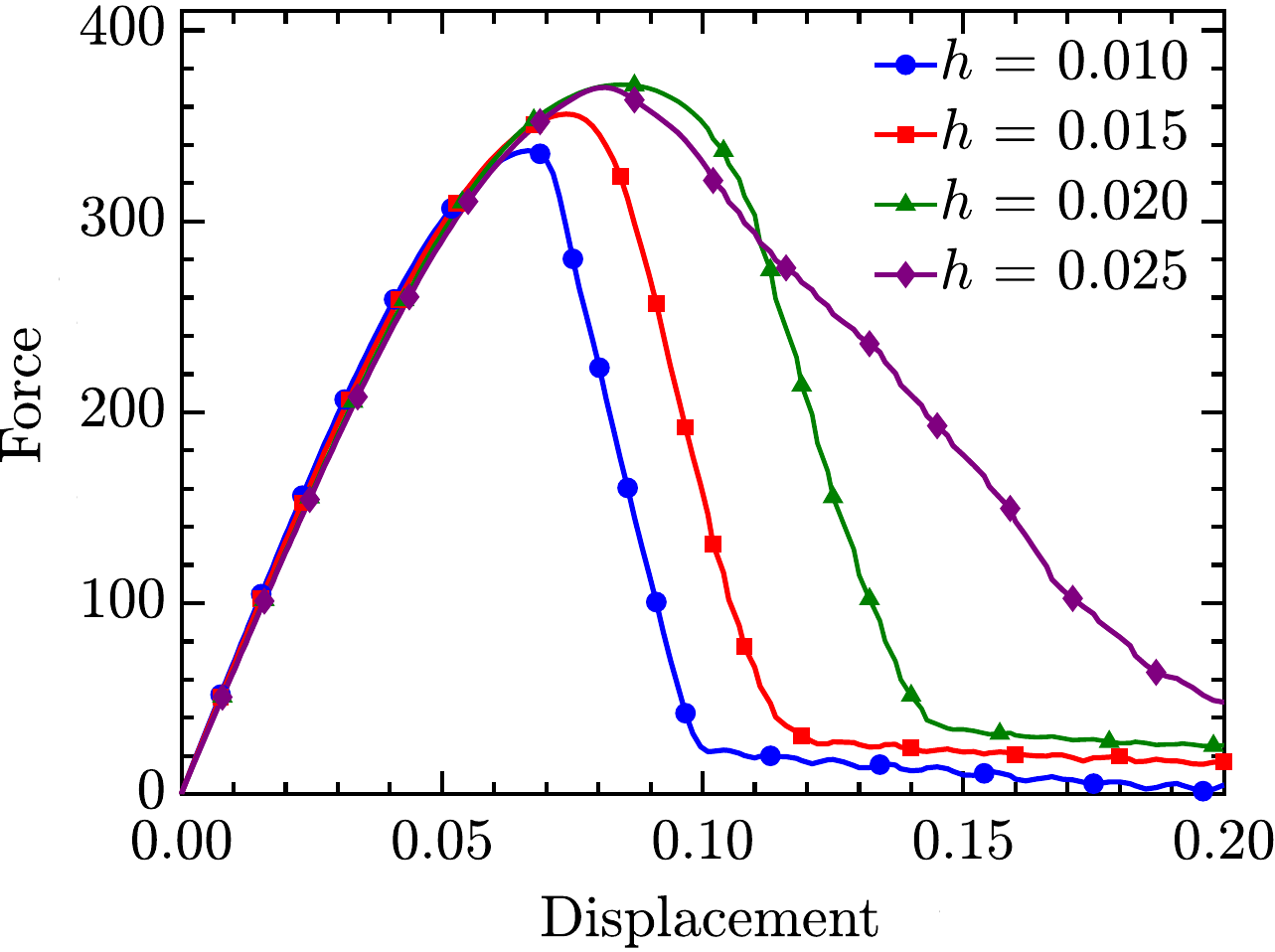}}
	\subfloat[Failure strain = 2.5\label{XX_FS_1c}]{\includegraphics[trim = 0 0 0 0, clip,height=3.5cm]{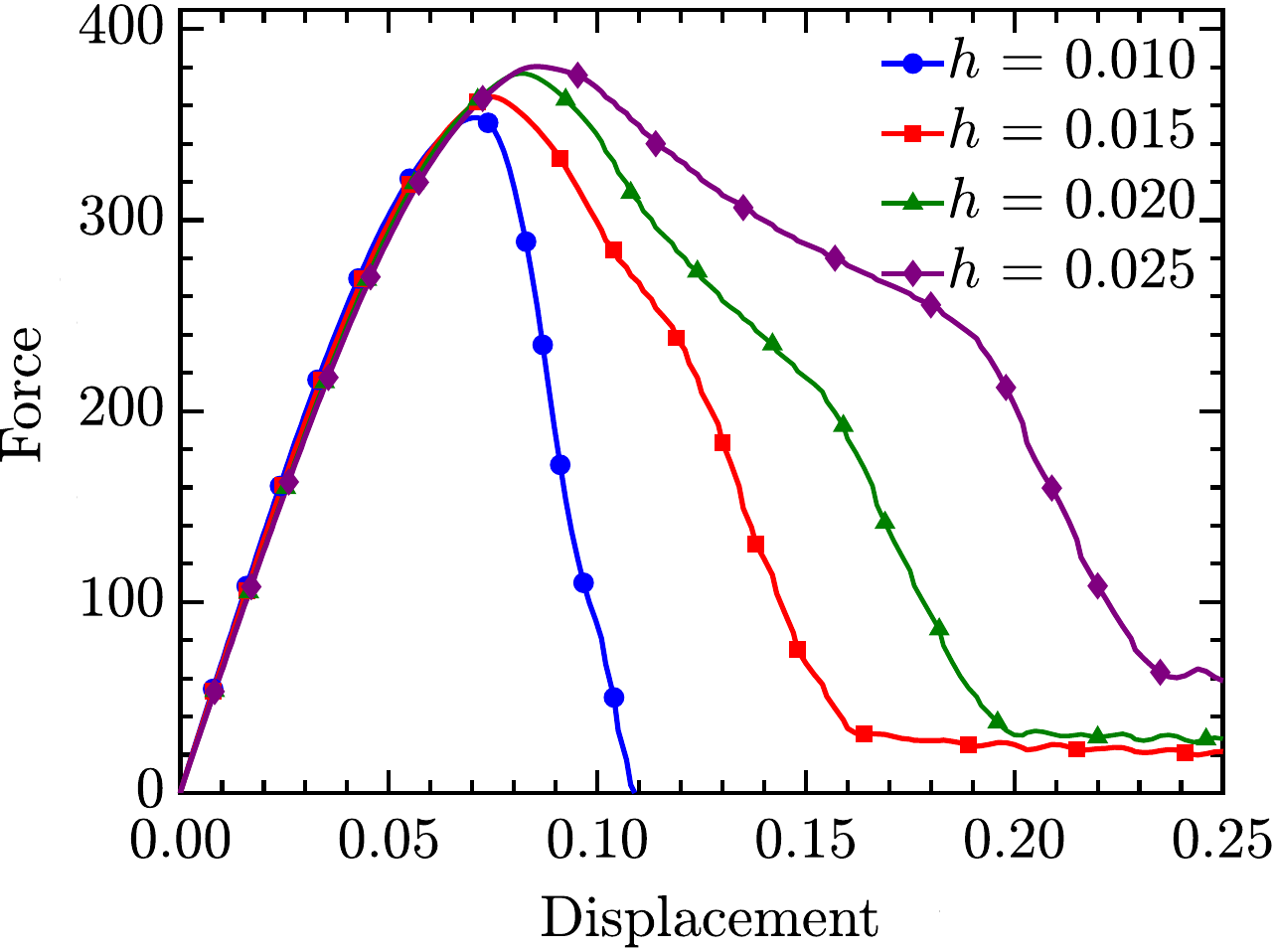}}
	\caption[$F$-$d$ curves of same RVE analysed with different mesh sizes h for different values of failure strains assigned.]{$F$-$d$ curves of same RVE analysed with different mesh sizes $h$ for different values of failure strains assigned as (a) $\varepsilon_f=1.5$, (b) $\varepsilon_f=2.0$ and (c) $\varepsilon_f=2.5$, All the samples are assigned a damage initiation strain of 0.125.}
	\label{XX_FS_1}
\end{figure}

 The RVE is analysed with uniform mesh sizes of 0.010 units, 0.015 units, 0.020 units, and 0.025 units. \color{black}Uniform mesh size facilitates a single value of the characteristic length and overall, it leads to less computational burden and easy implementation of the numerical formulation. \color{black}The crack path in each of these cases is shown in  \figurename~\ref{undeformed_XX_FS_1_5}. The $F$-$d$ curves corresponding to each of these analyses are shown in \figurename~\ref{XX_FS_1}\hyperref[XX_FS_1]{(a)}. 
\par The RVE studies are carried out by varying the failure strain ($\kappa_F=\varepsilon_f$) as 1.5, 2.0 and 2.5 and the $F$-$d$ curves obtained are shown in figures \ref{XX_FS_1}\hyperref[XX_FS_1]{(b)}, \ref{XX_FS_1}\hyperref[XX_FS_1]{(c)} respectively. 
\par It is observed from \figurename~\ref{XX_FS_1} that for a given value of failure strain, the larger the mesh size, the larger the area under $F$-$d$ curve, i.e., the larger the mesh size, the higher the amount of energy dissipated during damage. \figurename~\ref{undeformed_XX_FS_1_5} shows that:
\begin{enumerate}
	\item The width of the crack paths is seen to be exactly equal to the size of the mesh being used.
	\item The path in which damage propagates may vary with a change in mesh size.
\end{enumerate}
\par Hence, two potential factors for difference in material response of the same RVE are identified; (1). (crack width (or mesh size) and (2). path of damage propagation. An apparent attempt to address the challenge of spurious mesh sensitivity is by \cite{brekelmans-1995}. The proposed solution is based on the equalisation of dissipated energy in the actual case and the FE analysis. The energy dissipated per unit crack area $U_d$ is given by
\begin{equation}
	U_d = \lambda\int_0^{\varepsilon_f}\sigma d\varepsilon
	\label{eq_U_d}
\end{equation}
${\varepsilon_f}$ is the critical or failure strain. Since the material follows the linear damage evolution law (refer Figure \ref{quasibrittle_samplecurve}), Eq. (\ref{eq_U_d}) becomes
\begin{equation}
	U_d = \frac{E}{2}\lambda\varepsilon^0\varepsilon_f
	\label{U_d_basiceqn}
\end{equation}
where $E$ is the Young's modulus, and  $\varepsilon^0$ is the damage initiation strain.
\par As discussed before, \textbf{the bandwidth of the strain localisation zone, which is supposed to be a constant for a given RVE, is seen to take the value of mesh size} $h$ being used for FE analysis. This, in turn, changes the energy dissipated per unit crack area. Hence, the energy dissipated per unit area as predicted from the simulation $U_d^{\prime}$ becomes 
\begin{equation}
	U_d^{\prime} = \left(\frac{h}{\lambda}\right)U_d
\end{equation}
\par
\begin{figure}[H]
\centering
\includegraphics[trim = 0 0 0 0, clip, height=4.0cm]{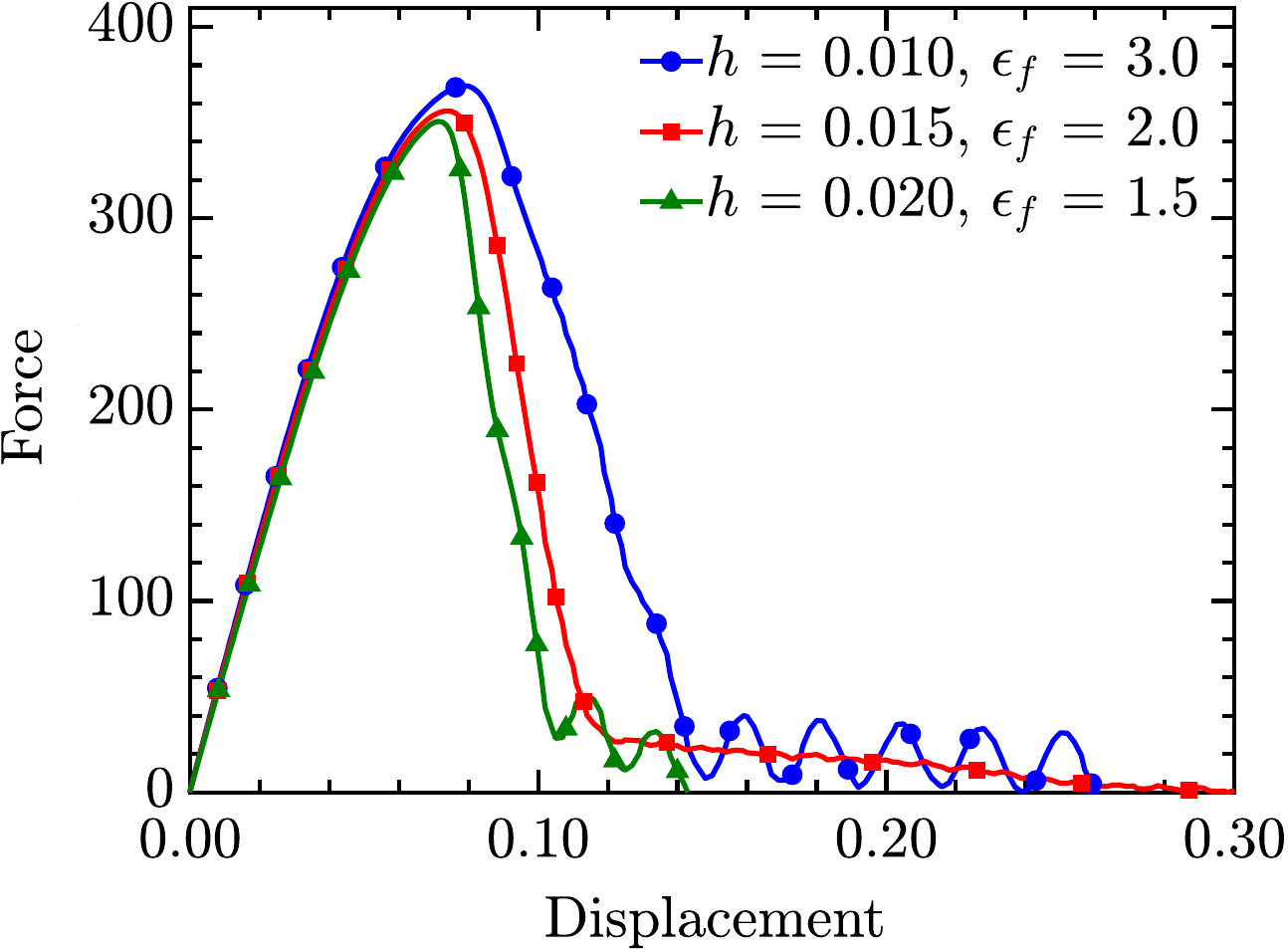}
\caption{$F$-$d$ curves of RVE samples analyzed by adjusting element size $h$ and failure strain $\varepsilon_f$ such that $h \varepsilon_f = 0.03$.}
\label{XY_ES_FS_3}
\end{figure}

 In order to equalize the value of $U_d^{\prime}$ to $U_d$, \cite{brekelmans-1995} proposed to change the value of critical strain $\varepsilon_f$ as 
\begin{equation}
	\varepsilon_{f,\ modified} = \left(\frac{\lambda}{h}\right)\varepsilon_{f,\ original}
	\label{epsilonalgorithmic}
\end{equation}
where $\varepsilon_{f,\ original}$ is the material's characteristic bandwidth of strain localization. And, $\varepsilon_{f,\ modified}$ is the value of failure strain to be considered while performing the simulation so that the total energy dissipated per unit crack area matches that from the original phenomenon. By this convention, an RVE, when analysed with two different element sizes, produce the same $F$-$d$ curve if the product of element size and failure strain is the same in both the cases (say, cases 1 and 2), i.e., if
\begin{equation}
	h_1\varepsilon_{f1}  = h_2\varepsilon_{f2} 
	\label{h1epsilon1}
\end{equation}
where subscripts 1 and 2 indicate that the terms correspond to cases 1 and 2, respectively. Following this criterion proposed by \cite{brekelmans-1995}, we performed simulations of which the results are presented in \figurename~\ref{XY_ES_FS_3}. The $F$-$d$ curve and the energy dissipated during damage still vary when the same RVE is analysed with different mesh sizes. This suggests that it is necessary to correct the criterion proposed in Eq. \eqref{h1epsilon1} to alleviate mesh sensitivity in RVE modelling. 
\subsection{A Technique to Alleviate Mesh Sensitivity in RVE Modeling}
\label{A Technique to Alleviate Mesh Sensitivity in RVE Modeling}
\begin{figure}[H]
	\centering
	\includegraphics[width=0.35\textwidth]{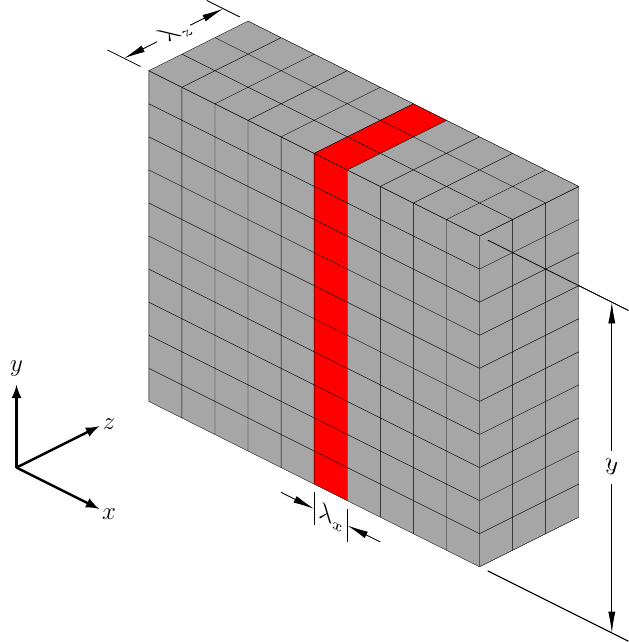}
		\caption{Diagram illustrating the bandwidth of strain localisation.}
	\label{lambdaxlambday}
\end{figure}

Let us consider a 3D domain fixed in the $xx$ direction on the left face, and a load in the form of displacement be applied on the right face. Let $A$ be the area of the crack surfaces. The total dissipated energy is given by
\begin{equation}
	(U_d)_{total} = U_d A = \frac{1}{2}E\varepsilon^0\varepsilon_f\lambda_x\lambda_zy 
	\label{U_d_total}
\end{equation}
where $\lambda_x$ is the band width of strain localization zone in the $x$ direction (\figurename~\ref{lambdaxlambday}). A characteristic bandwidth of strain localisation $\lambda_{modified}$ could be assigned to the domain as the geometric mean of $\lambda_x$ and $\lambda_z$ as
\begin{equation}
	\lambda_{modified} = \sqrt{\lambda_x\lambda_z}
	\label{squarerootxy}
\end{equation}
Substituting (\ref{squarerootxy}) in (\ref{U_d_total}),
\begin{equation}
	(U_d)_{total} = \frac{1}{2}E\varepsilon^0\varepsilon_f(\lambda_{modified})^2y 
	\label{U_d_total_new}
\end{equation}
\par A 2D plane stress case can be considered a special case of the 3D, where the dimension in its out-of-plane direction is 1 unit. In such a case, $\lambda_z = 1$, and hence,
\begin{equation}
	\lambda_{modified} = \sqrt{\lambda_x}
\end{equation} 
Since the bandwidth of the strain localisation zone ${\lambda_x}$ as observed from the output from the FEA package, is equal to the element size $h$ being used,
\begin{equation}
	\lambda_{modified} = \sqrt{h}
\end{equation} 
The energy dissipated per unit area, as observed from the FE analyses, becomes
\begin{equation}
	U_d^{\prime} = \frac{E}{2}\lambda\varepsilon^0\varepsilon_f = \frac{E}{2}\sqrt{h}\varepsilon^0\varepsilon_f
	\label{U_d_new_basiceqn}
\end{equation}
The energy dissipated during crack formation is directly proportional to $\sqrt{h}\varepsilon_f$. Hence, an RVE, when analysed with two different element sizes, produce the same $F$-$d$ curve if the product of failure strain and the square root of element size is the same in both cases (say, cases 1 and 2), i.e., if
\begin{equation}
	\sqrt{h_1}\varepsilon_{f1} = \sqrt{h_2}\varepsilon_{f2} 
	\label{new_technique_eqn}
\end{equation}
Eq. \eqref{new_technique_eqn} also suggests that the energy dissipated during damage, as well as the $F$-$d$ curves as observed from FE analysis, will match with those obtained from the actual phenomenon if
	\begin{equation}
		(\varepsilon_f)_{modified} = \sqrt{\frac{\lambda}{h}}(\varepsilon_f)_{original}
		\label{new_technique_eqn_experimental}
	\end{equation}
\par
\begin{figure}[]
	\centering
	\subfloat[$\sqrt{h} \varepsilon_f = 0.282$]{\includegraphics[trim = 0 0 0 0, clip, height=3.65cm]{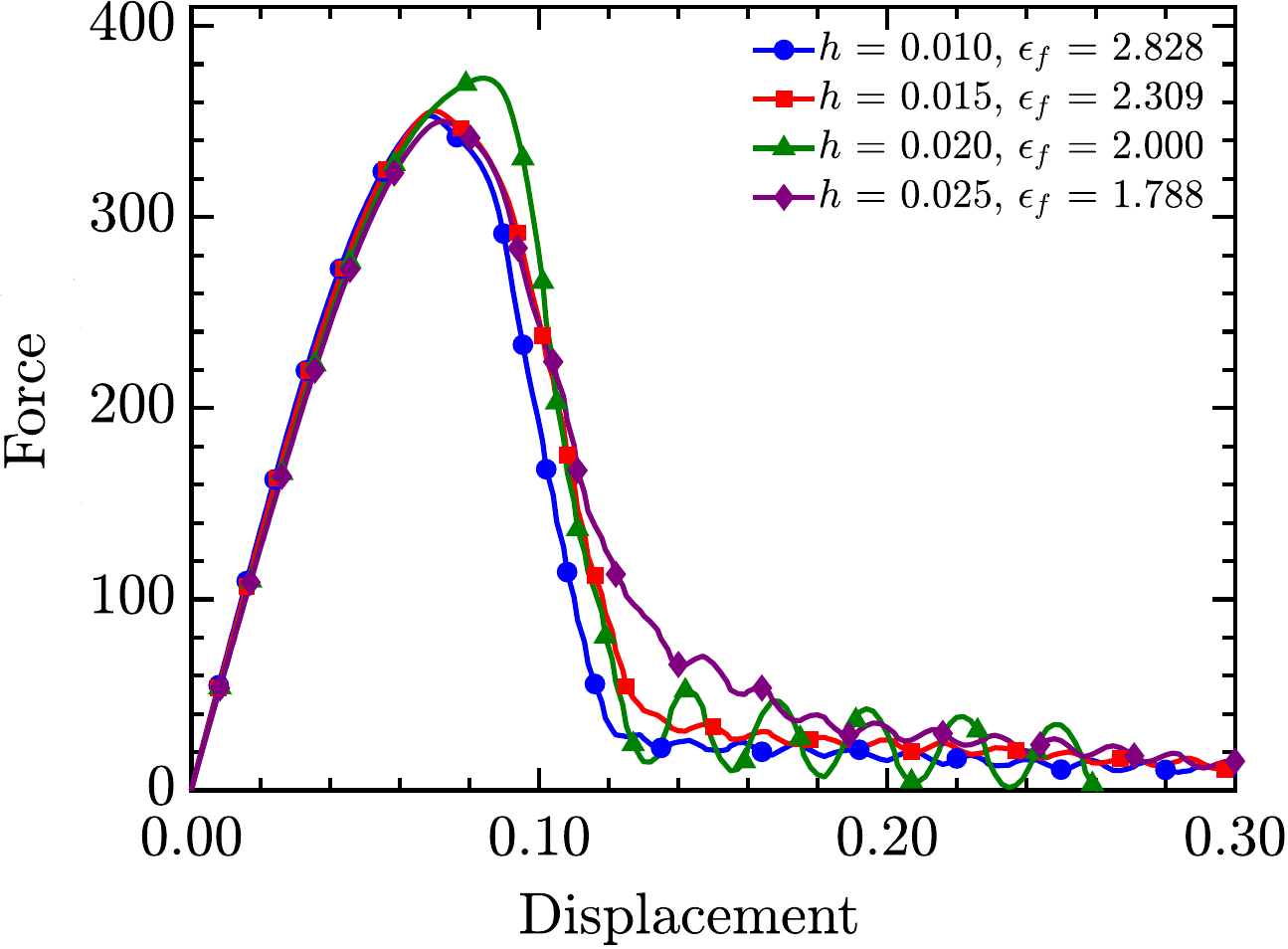}}
	\subfloat[$\sqrt{h} \varepsilon_f = 0.346$]{\includegraphics[trim = 0 0 0 0, clip, height=3.65cm]{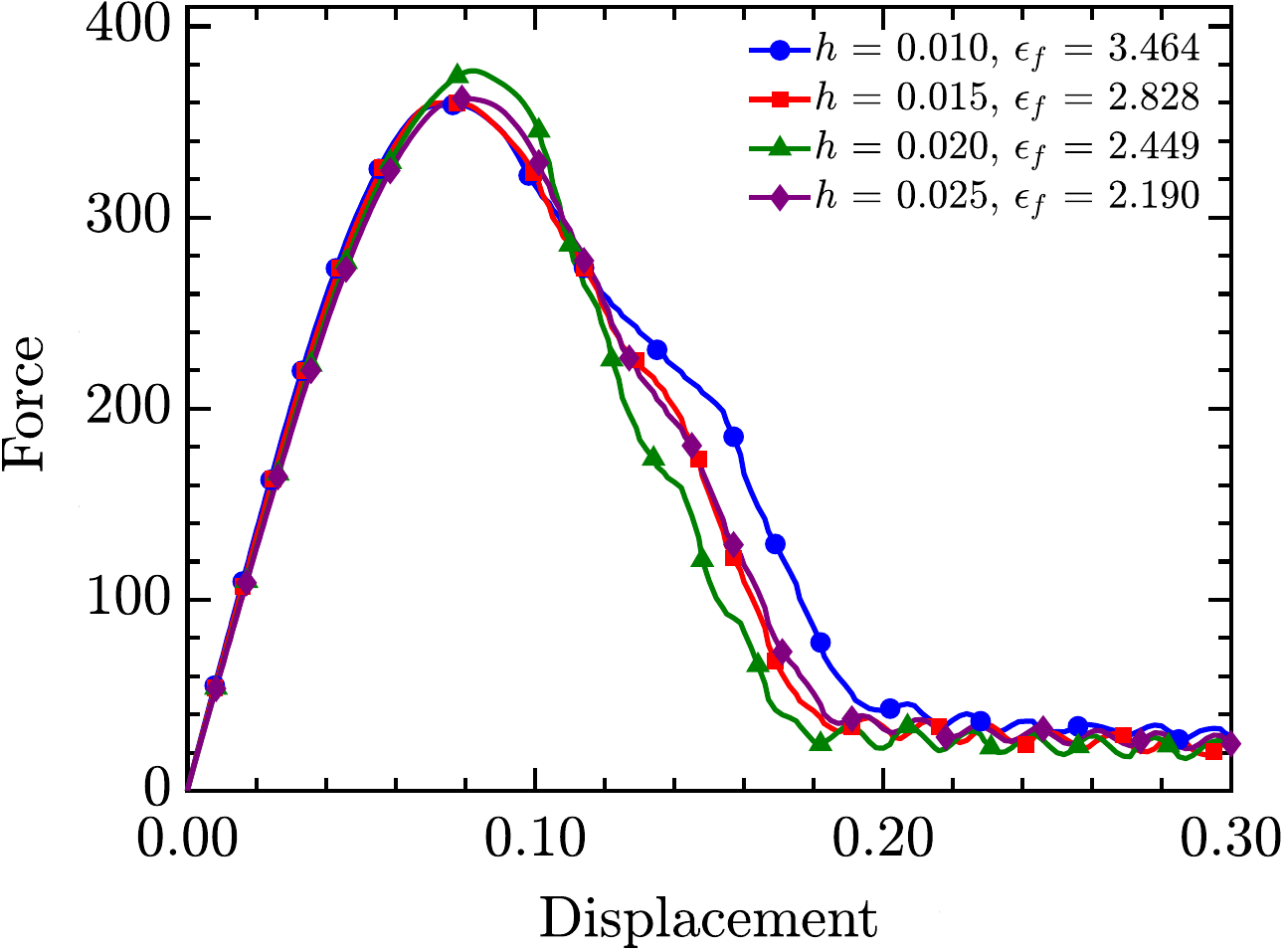}}
	\subfloat[$\sqrt{h} \varepsilon_f = 0.400$]{\includegraphics[trim = 0 0 0 0, clip, height=3.65cm]{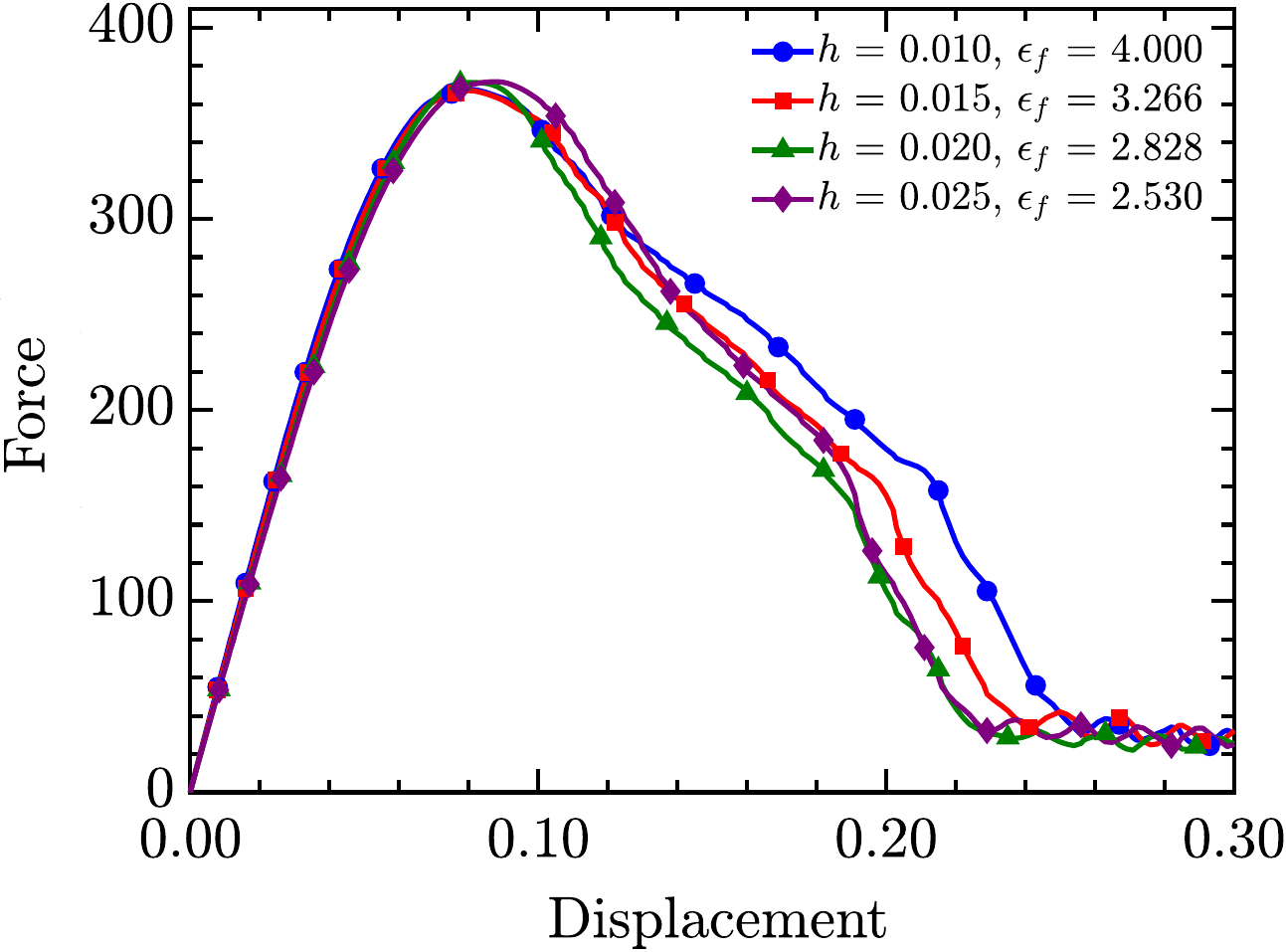}}
	\caption{$F$-$d$ curves with the element sizes ($h$) and failure strains ($\varepsilon_{f}$) adjusted according to the technique proposed in Eq. (\ref{new_technique_eqn})}
	\label{Matching_curve_YY}
\end{figure}

 The RVE used for the previous analyses shown in this chapter is analysed by adjusting the element sizes and failure strains such that they follow the criterion above put forward by (\ref{new_technique_eqn}) (see \figurename~\ref{Matching_curve_YY}). The $F$-$d$ curves within each subplot in Figure \ref{Matching_curve_YY} are similar to a good extent. This verifies the newly proposed technique to alleviate the mesh size sensitivity in RVE modelling.
\section{RVE Size Dependent Softening Response}
\label{RVE Size Dependent Softening Response}
This section investigates how the material response varies with RVE size at different fibre volume fractions. This section also proposes a set of modified periodic boundary conditions (MPBCs) to converge the observed material response from FE analysis at a higher frequency, even at smaller RVE sizes, to that of the actual one.
\begin{figure}[H]
	\centering
	\subfloat[]{\includegraphics[trim = 0 0 0 0, clip, width=6cm]{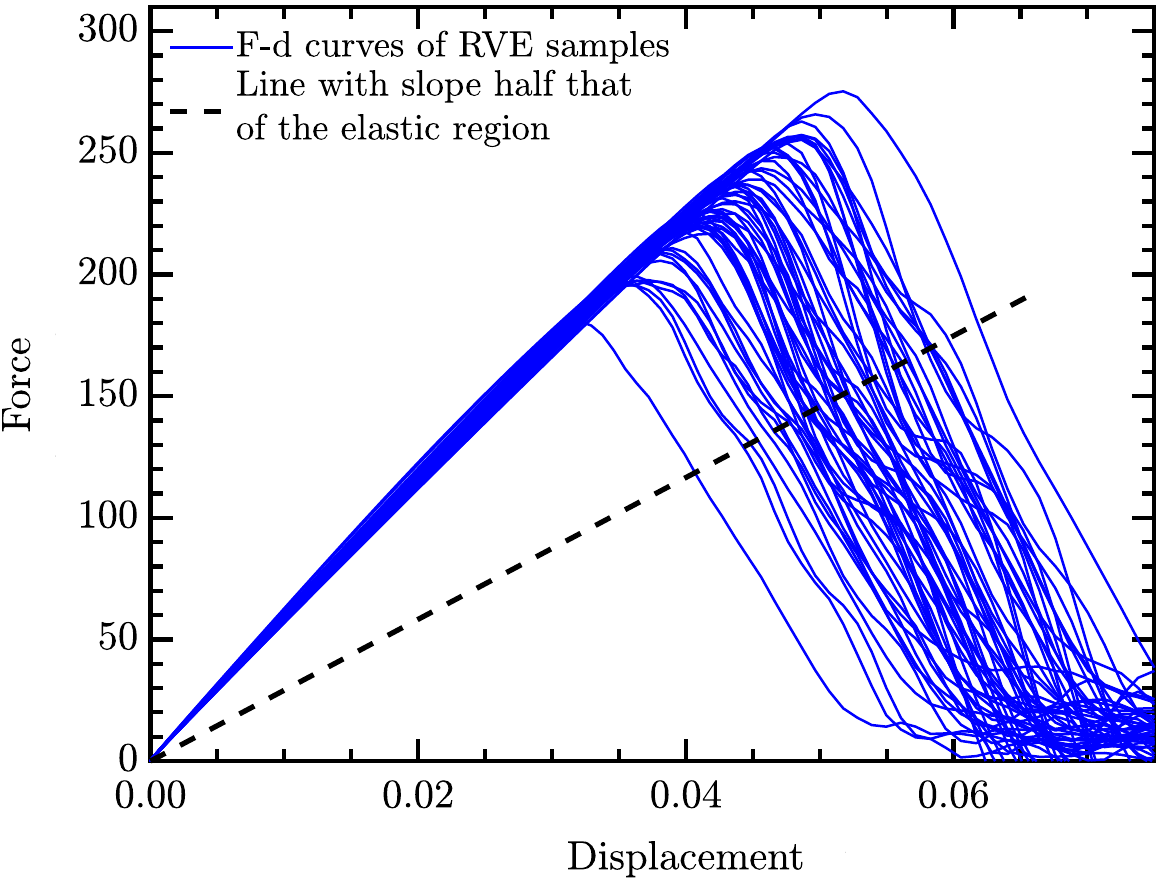}} 
	\subfloat[]{\includegraphics[trim = 0 0 0 0, clip, width=6.2cm]{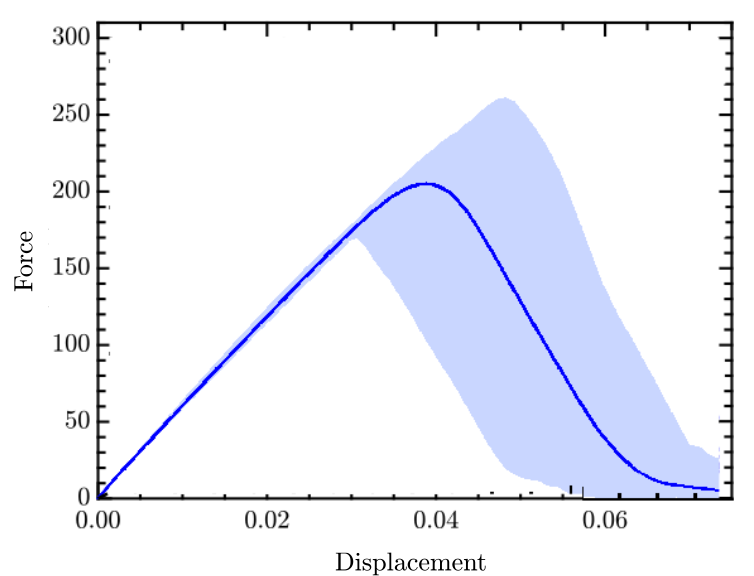}}
	\caption{(a) $F$-$d$ curves of 60 RVE samples having different fibre sizes and a total fibre volume fraction of 0.30. (b) The mean $F$-$d$ curve with the region between these $F$-$d$ curves in (a) is shaded in grey. The dashed line corresponds to half the slope of the elastic region.}
	\label{postdamagevariation}
\end{figure}
\par The window size of the RVEs considered in this study is always 1 unit $\times$ 1 unit. But, the number of fibres considered inside the RVEs is 15, 30, 40, and 50. An RVE with 15 fibres may be deemed to be zoomed in to a reasonable degree, while that with 50 fibres may be observed more zoomed out. When we consider an RVE with 15 fibres, the actual size of the RVE under focus is smaller than when an RVE of 30 fibres is assessed, and so on. 
\begin{figure}[H]
	\centering
	\includegraphics[width=\textwidth]{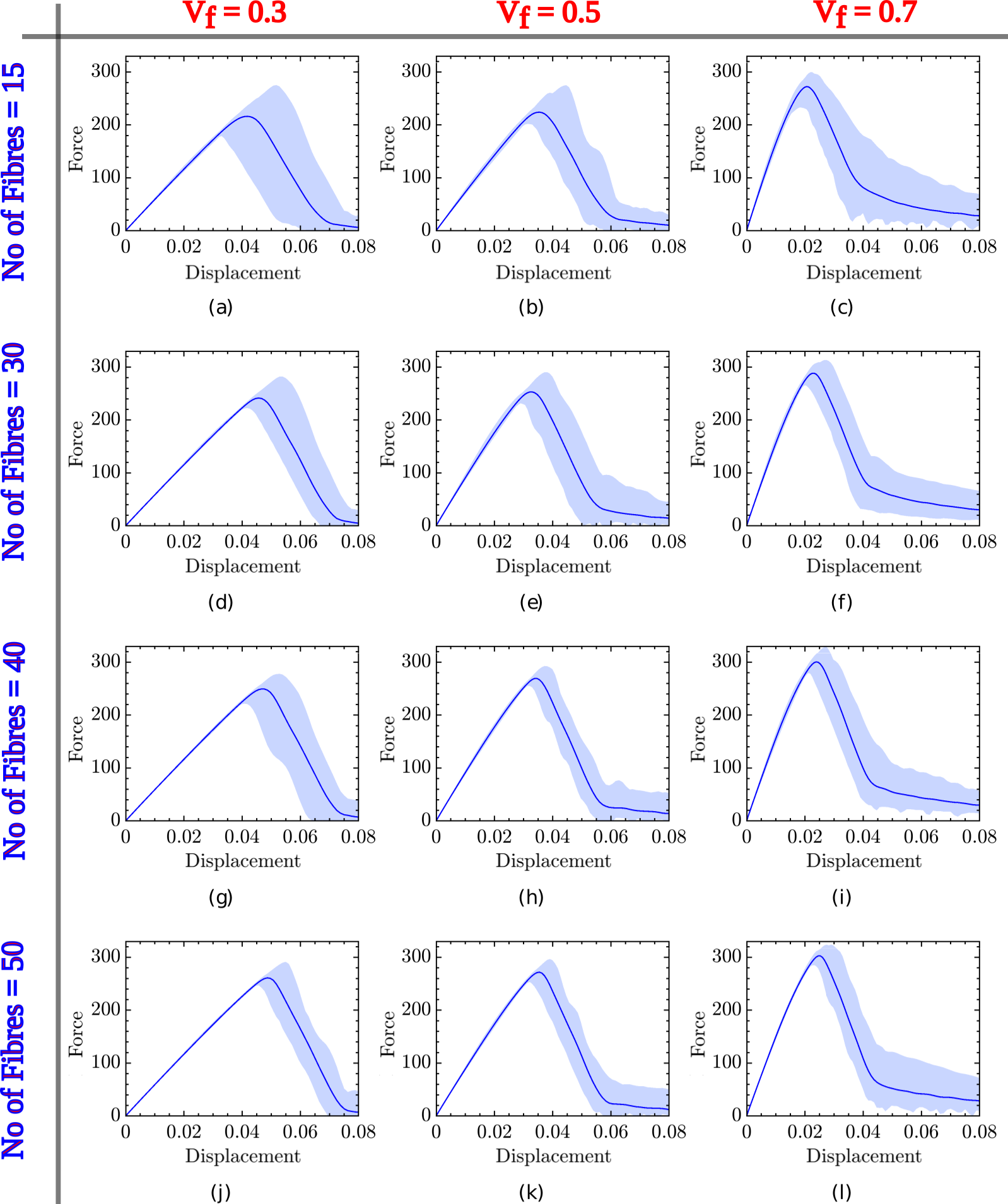}
	\caption{$F$-$d$ curves for RVE samples with a number of fibres as 50, and fibre volume fraction ${V_f}$ as (a) 0.30, (b) 0.50, (c) 0.70 (samples analysed with the application of displacement PBCs). Initially, the $F$-$d$ curves of the 60 samples considered for each volume fraction are plotted. Then, the region between these $F$-$d$ curves in each subplot is shaded in grey. A mean $F$-$d$ curve, represented by the thin blue curve, is obtained by averaging the force values of the 60 samples at each load step.}
	\label{XXDPBC50}
\end{figure}
\par
\begin{figure}[H]
	\centering
	\includegraphics[width=\textwidth]{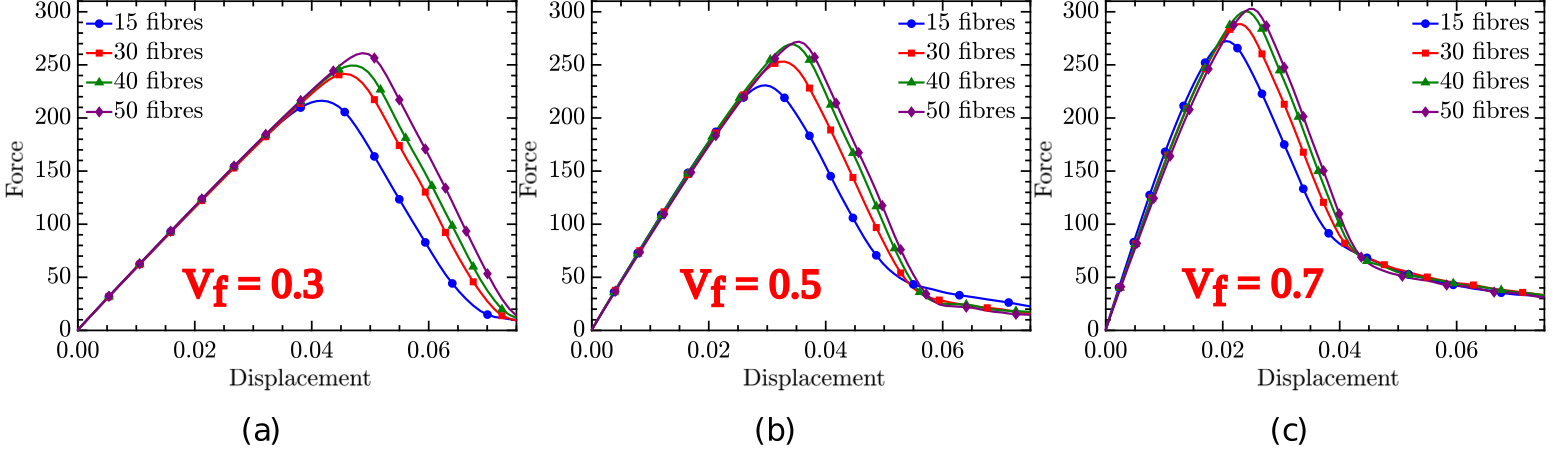}
	\caption{Average $F$-$d$ curves at different RVE sizes and fibre volume fraction as (a) 0.30, (b) 0.50, and (c) 0.70 (samples analysed with the application of displacement PBCs).}
	\label{meanvalcurvedpbc}
\end{figure}

 It is observed that the $F$-$d$ curves are almost coinciding in the elastic region, and noticeable dispersion occurs among them once the damage is initiated (\figurename~\ref{postdamagevariation}(a)). For each combination of fibre volume fraction and fibre size, a mean $F$-$d$ curve is obtained (\figurename~\ref{postdamagevariation}(b)). Sixty RVEs each, having 15, 30, 40 and 50 circular inhomogeneities, are considered for $V_f$s 0.30, 0.50 and 0.70. (\figurename~\ref{XXDPBC50}). The mean $F$-$d$ curves for the four different RVE sizes considered for a given fibre volume fraction are plotted together (\figurename~\ref{meanvalcurvedpbc}), and the dispersion among these curves are used as a measure to quantify the effect of RVE size on the material response at that particular fibre volume fraction.
\par It is observed that with the increase in fibre volume fraction, the peak force (and hence, the peak stress) in the samples increases while the failure occurs at a smaller displacement value. It is also observed that the individual curves in each of these $F$-$d$ plots are almost coinciding in the elastic region, but they show dispersion post-damage initiation. This can be attributed to the fact that the elastic properties depend on the initial microstructure. In contrast, the fracture properties depend on the microstructure induced by the network of developing cracks \citep{pelissou-2009}. The localisation of damage leads to a variation of material response after damage initialisation among different RVE samples. However, their volume fractions and several fibres (or RVE size) are the same. The material response post-damage initiation is seen to vary considerably with the size of the RVE being used (\figurename~\ref{meanvalcurvedpbc}). 
\par It can be observed that the peak force (and hence, the peak stress) increases with an increasing number of fibres. In all the investigated volume fractions, RVEs with 15 inhomogeneities registered the lowest peak stress, and those with 50 fibres registered the highest peak stress. Also, the strain at which the peak stress occurs increases with the number of fibres. The ability of the RVEs of a given fibre volume fraction to withstand a higher strain, with the increase in the number of fibres, may be attributed to the fact that the higher the number of fibres for a given volume fraction, the more evenly distributed the fibres are. Speaking in terms of elements in the domain in the FEA model, the higher the number of fibres for a given volume fraction, the lesser the accumulation of elements assigned to the material properties of the fibre.
\par Post damage dispersion of $F$-$d$ curves is quantified using the concept of standard deviation. The average slope of the elastic region of these curves is found. And, a line having a slope half of this average slope is drawn (\figurename \ref{postdamagevariation}(a)). This line would cut each of the $F$-$d$ curves in the region after damage initiation. The displacement values corresponding to the point of intersection of each of the $F$-$d$ curves and this line are noted. The standard deviation of these displacement values is obtained using Eq. (\ref{eqnstddev}). It is considered a measure of post-damage dispersion of $F$-$d$ curves for that particular combination of fibre volume fraction and RVE size.
\begin{equation}
	SD = \sqrt{\frac{\Sigma(x_{i} - \mu)^{2}}{n}}
	\label{eqnstddev}
\end{equation}
where $SD$ is the population standard deviation, $n$ is the size of the population, $x_i$ represents each value from the population, and $\mu$ is the population mean. 
\par
\begin{figure}[H]
	\centering
	\includegraphics[trim = 0 0 0 0, clip, width = 7.0cm]{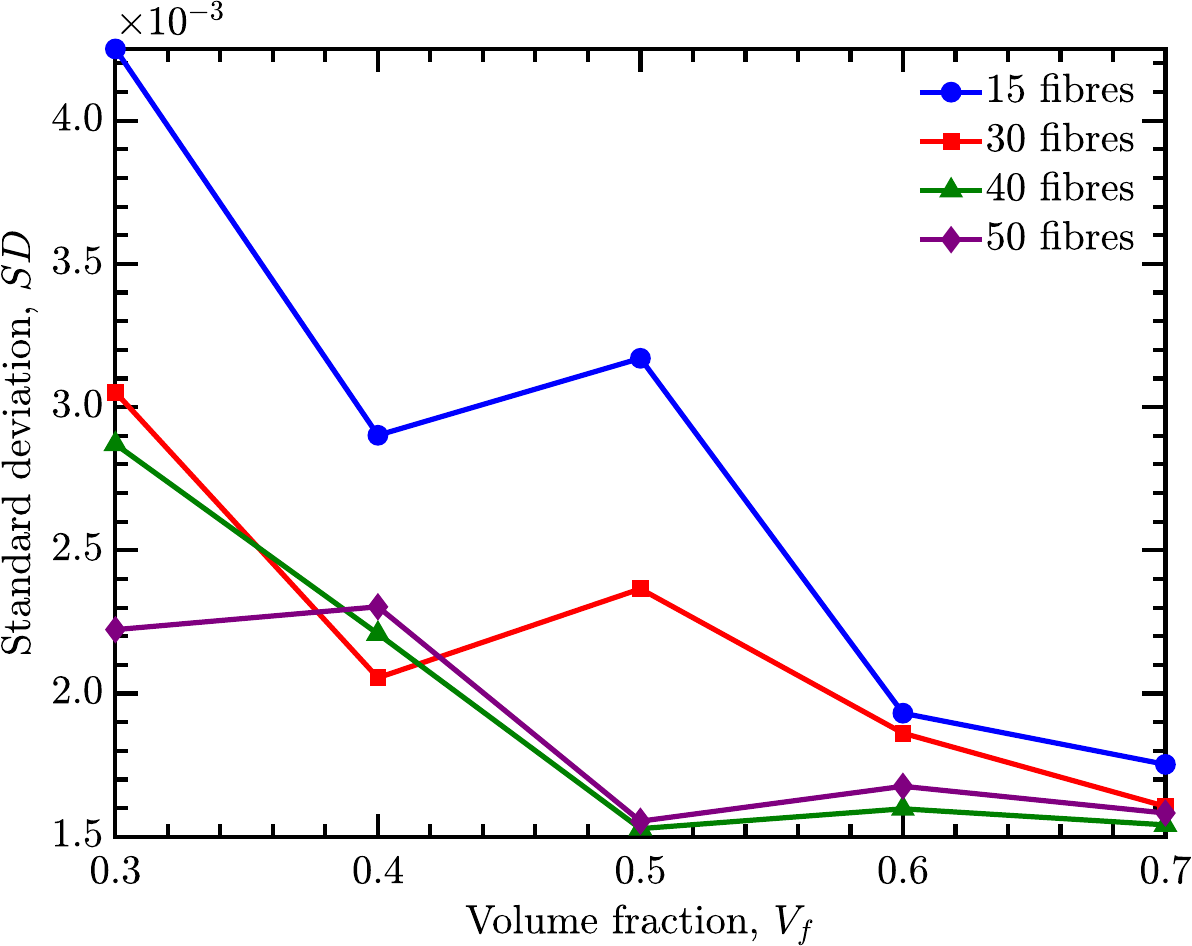}
	\caption{Standard deviation of the post damage dispersion of $F$-$d$ curves as a function of volume fraction at the four RVE sizes considered with a different number of fibres.}
	\label{Comparison_sd1}
\end{figure}

 The standard deviation of $F$-$d$ curves after damage initiation at different fibre volume fractions, at the RVE sizes considered, i.e., from the smallest sets of RVEs having 15 inhomogeneities to the largest ones having 50 inhomogeneities, are given in \tablename~\ref{Table_comparison_dpbc_mpbc_empbc}. Based on this data, the trend in post-damage dispersion of material response of RVE samples, with fibre volume fraction, is plotted in \figurename~\ref{Comparison_sd1}. Results indicate that post-damage dispersion of $F$-$d$ curves decreases with increased fibre volume fraction. Hence, at a given RVE size, the computationally observed material response of RVEs will be precise for higher fibre volume fractions compared to those of RVEs with smaller fibre volume fractions. The results also indicate that as the number of fibres in the RVEs increases, i.e., as the size of the RVE increases, the dispersion among the $F$-$d$ curves of RVE samples for a given volume fraction decreases, i.e., the material response from different RVEs having the same fibre volume fraction becomes more precise.
\subsection{Modified Periodic Boundary Conditions}
\label{Modified Periodic Boundary Conditions}
\par The PBCs usually prescribed in RVE analysis is displacement PBCs, which ensures only the periodicity of displacement on the RVE boundaries. A set of modified PBCs (MPBC) is being proposed. In addition to ensuring displacement periodicity, it ensures strain periodicity of a region of thickness equal to the mesh size on the boundaries. This is based on the hypothesis that ensuring strain periodicity on the RVE boundaries assists in reducing the effect of RVE size in microscale damage analysis. 
\par
\begin{figure}[H]
	\centering
	\includegraphics[width=0.5\textwidth]{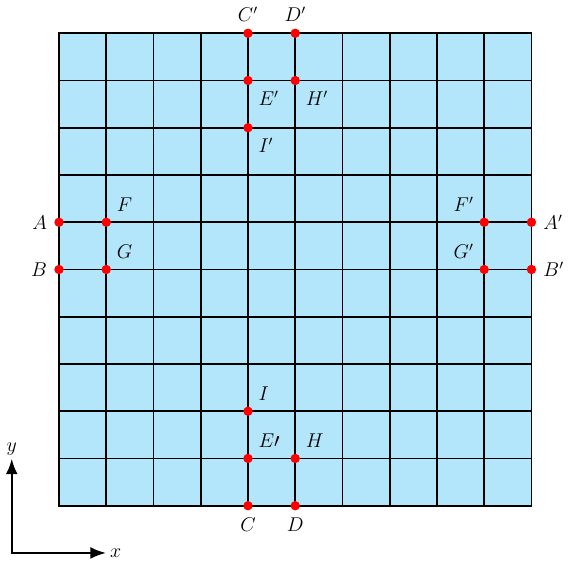}
	\caption{A sample mesh to demonstrate the application of modified PBCs over an extended region near the boundaries, in $xx$ loading.}
	\label{EPBC_illustration_mesh}
\end{figure}

 Consider a domain meshed as in \figurename~\ref{EPBC_illustration_mesh}. Suppose that the same is to be loaded as in $xx$ loading condition and that MPBCs are to be applied in a region having a width greater than the usual width by a unit equal to mesh size. All the constraints applied for displacement PBCs are applied, and additionally, the constraints need to be prescribed to equalise strain in the layer penultimate to the boundary. The general form of MPBCs in such a case is as given in Eq. (\ref{EPBC_ref}).\newline
On the boundary:
\begin{subequations}
	\allowdisplaybreaks[1]
	\label{EPBC_ref}
	\begin{align}
		\label{epbc_eq_a}
		u_{y}^{A} &= {u_{y}^{A^{\prime}}}\\
		\label{epbc_eq_b}
		u_{y}^{A} - u_{y}^{B} &= u_{y}^{A^{\prime}} - {u_{y}^{B^{\prime}}}\\
		\label{epbc_eq_c}
		u_{x}^{C} &= u_{x}^{C^{\prime}}\\
		\label{epbc_eq_d}
		u_{x}^{C} - u_{x}^{D} &= u_{x}^{C^{\prime}} - {u_{x}^{D^{\prime}}}\\
		\label{epbc_eq_e}
		u_{y}^{C} &= {u_{y}^{C^{\prime}}}\\
		\label{epbc_eq_f}
		u_{y}^{C} - u_{y}^{E} &= u_{y}^{E^{\prime}} - {u_{y}^{C^{\prime}}}
		\intertext{To ensure strain periodicity on the layer next to the boundary:}
		\label{epbc_eq_g}
		u_{y}^{F} - u_{y}^{G} &= u_{y}^{F^{\prime}} - {u_{y}^{G^{\prime}}}\\
		\label{epbc_eq_h}
		u_{x}^{E} - u_{x}^{H} &= u_{x}^{E^{\prime}} - {u_{x}^{H^{\prime}}}\\
		\label{epbc_eq_i}
		u_{y}^{E} - u_{y}^{I} &= u_{y}^{I^{\prime}} - {u_{y}^{E^{\prime}}}
	\end{align}
\end{subequations}
\par
\begin{figure}[H]
	\centering
	\subfloat{\includegraphics[width=5.0cm]{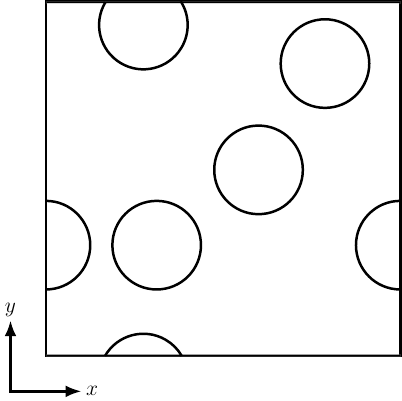}}
	\subfloat{\includegraphics[width=5.0cm]{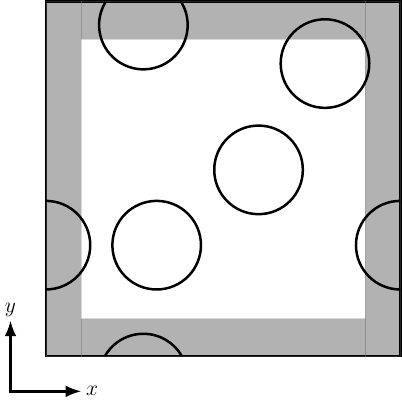}}
	\caption[Diagram illustrating the application of MPBCs, in a domain of size 1 unit $\times$ 1 unit discretised into 40,000 finite elements, (a) to a region of width equal to the mesh size on the boundaries, and (b) to a region of width approximately equal to ten times the mesh size, near the boundaries.]{Diagram illustrating the application of MPBCs, in a domain of size 1 unit $\times$ 1 unit discretised into 40,000 finite elements, (a) to a region of width equal to the mesh size on the boundaries, and (b)to a region of width approximately equal to ten times the mesh size, near the boundaries. The shaded regions are prescribed MPBCs. Since the $xx$ loading condition is used in this section of the study, strain in the $y$ direction is periodic on the top and bottom regions and the left and right regions. Strain in the $x$ direction is periodic on the top and bottom regions. Displacement in the $y$ direction is periodic on the top and bottom boundaries and the left and right boundaries. Displacement in the $x$ direction is periodic on the top and bottom boundaries.}
	\label{width_MPBC_EPBC}
\end{figure}

 Following this framework, MPBCs could be applied to a wider region near the RVE boundaries, of which the width is a multiple of the mesh size being used. As mentioned, the RVEs being analysed are of unit size, discretised into 40,000 finite elements, and the mesh size is 0.005 units. There are 200 elements along its length. MPBCs are applied to a region having a width of 10 elements near each boundary/edge of the RVE (\figurename~\ref{width_MPBC_EPBC}). 
\def\centerarc[#1](#2)(#3:#4:#5)
{ \draw[#1] ($(#2)+({#5*cos(#3)},{#5*sin(#3)})$) arc (#3:#4:#5); }
\tikzset{
	semi thick/.style={line width=1.4pt}
}
To verify the hypothesis that ensuring strain periodicity on the boundaries of the RVE assists in reducing the effect of RVE size in RVE modelling, analyses are performed on the same RVE samples utilised at the beginning of this section, with the application of MPBCs in place of the usually prescribed displacement PBCs. Except for the choice of PBCs, all other factors, including loading condition, damage initiation criteria, and damage evolution law, are kept the same. On the constrained edge and the edge on which the load is applied, modified PBCs are applied in a direction other than that in which the load is applied. On the other two edges, MPBCs are applied in both directions. 
\begin{table}[H]
	\centering
	\caption[Dispersion among $F$-$d$ curves of samples of different RVE sizes, at each of the fibre volume fractions, considered.]{Dispersion among $F$-$d$ curves of samples of different RVE sizes, at each of the fibre volume fractions considered. Standard deviations from the three sets of analysis, i.e., from the analysis of samples prescribed only with displacement PBCs (DPBC), modified PBCs (MPBC), and modified PBCs at an extended region near the boundary (MPBC), are tabulated. The dispersion at three regions of the $F$-$d$ curves are considered, i.e., at 0.25 times, 0.50 times, and 0.75 times the slope of the elastic region.\label{Table_comparison_dpbc_mpbc_empbc}}
		\begin{tabular}{@{}cccccccccc@{}}
			\toprule
			\multirow{3}{*}{\begin{tabular}[c]{@{}c@{}}Fibre\\Volume\\ fraction, $V_f$\end{tabular}} & \multicolumn{9}{c}{Standard deviation, $SD$ ($\times 10^{-3}$) \si{m}} \\ \cmidrule(l){2-10} & \multicolumn{2}{c}{0.25$\times$slope} & \multicolumn{2}{c}{0.50$\times$slope} & \multicolumn{2}{c}{0.75$\times$slope} \\\cmidrule(lr){2-4}\cmidrule(lr){5-7}\cmidrule(l){8-10}
			& DPBC  & EMPBC & DPBC  & EMPBC & DPBC  & EMPBC \\ \midrule
			0.30 & 2.616  & 2.327 & 2.668 & 2.408 & 2.680  & 2.442\\
			0.40 & 2.575  & 2.408 & 2.465  & 2.359 & 2.415  & 2.355\\
			0.50 & 1.854 &  1.841 & 1.913  & 1.899 & 2.073  & 1.991\\
			0.60  & 1.872  & 1.794 & 1.840  & 1.780 & 1.787  & 1.731\\
			0.70 & 1.331  & 1.224 & 1.374  & 1.288 & 1.280  & 1.170\\ \bottomrule
		\end{tabular}
\end{table}
\begin{figure}[H]
	\centering
	\includegraphics[trim = 0 0 0 0, clip, width = 7.0cm]{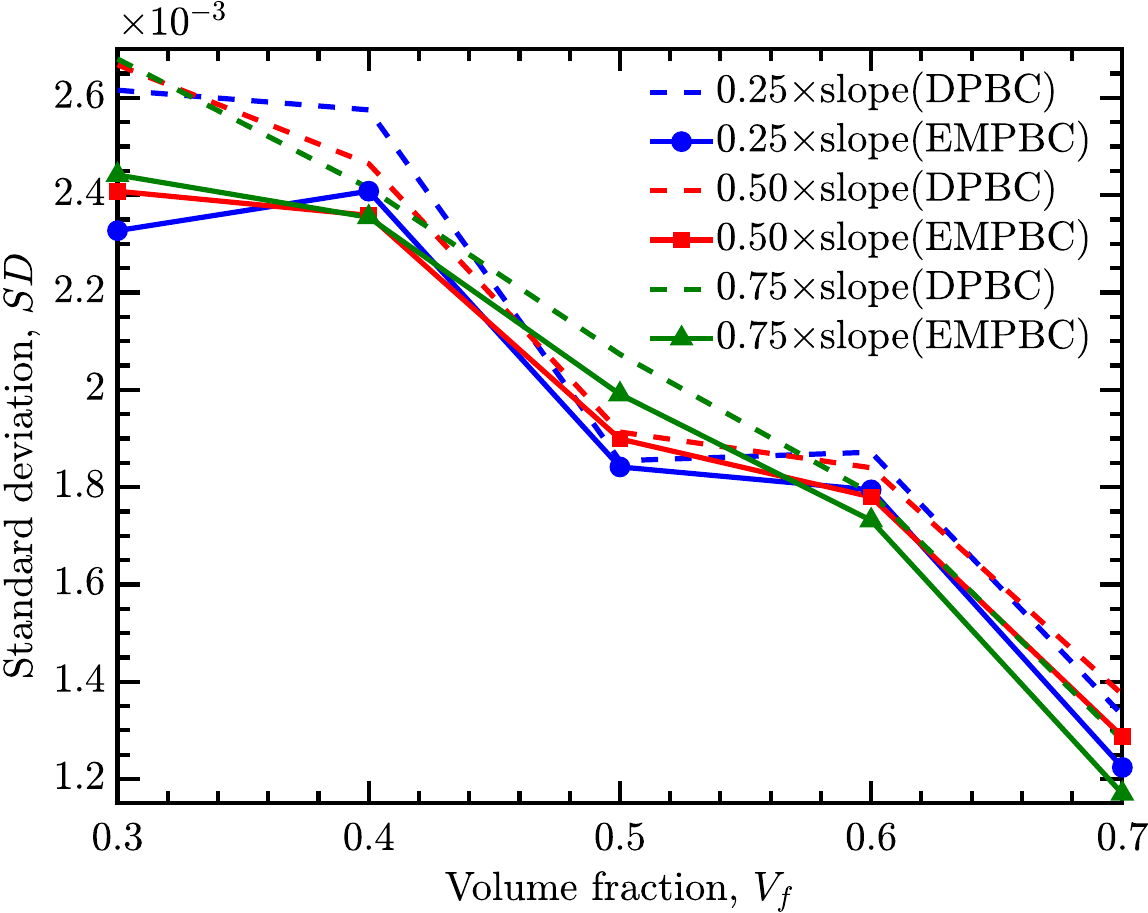}
	\caption[Dispersion among $F$-$d$ curves corresponding to the four RVE sizes considered.]{Dispersion among $F$-$d$ curves corresponding to the four RVE sizes considered. Each data point in the plot quantifies the post-damage dispersion at a particular region (at either 0.25 times, 0.50 times or 0.75 times the slope of the elastic region), among the four $F$-$d$ curves (each of these $F$-$d$ curves are obtained by averaging the $F$-$d$ curves of 60 samples considered in that combination of volume fraction and RVE size) when analysed by prescribing a particular PBC (either displacement PBC (DPBC) or MPBC over an extended region across the boundary (MPBC)).}
	\label{DPBC_EPBC_Size_Effect}
\end{figure}
\par From \figurename~\ref{DPBC_EPBC_Size_Effect}, the following observations can be made:
\begin{enumerate}
	\item{The higher the fibre volume fraction, the lesser the dispersion among the average $F$-$d$ curves corresponding to different RVE sizes. This trend remains the same at all three regions of the $F$-$d$ curves considered; hence, it is logical to assume that this observation will remain valid throughout the post-damage region of the $F$-$d$ curves. It can be concluded that the effect of RVE size on material response is lesser at higher fibre volume fractions.}
	\item{The standard deviation vs volume fraction curves corresponding to the analyses done with displacement PBC, and those corresponding to the analyses done with MPBC, at all three regions of the $F$-$d$ curves considered, are almost coinciding. This means that the application of MPBCs has not exhibited a noticeable effect in reducing RVE size sensitivity.}
\end{enumerate}
\par The standard deviation of $F$-$d$ curves corresponding to different RVE sizes of a given fibre volume fraction is noticeably lesser when the samples are analysed by prescribing MPBCs on a wider region across the RVE boundaries (\tablename~\ref{Table_comparison_dpbc_mpbc_empbc}, \figurename~\ref{DPBC_EPBC_Size_Effect}). It can be inferred that MPBCs have successfully demonstrated their capability to reduce RVE size sensitivity in RVE modelling.
\section{Hetrogeneity Distribution Dependent Softening Response}
\label{Hetrogeneity Distribution Dependent Softening Response}
The previous sections indicated that the material response of RVEs depends upon their position of heterogeneities. This section identifies two key factors influencing the possibility of damage initiation in a location within the RVE and quantifies their relative contribution in deciding the extent to which the location favours damage initiation. The findings from this section are significant due to their potential to bring in the following advances:
\begin{enumerate}
\item{Insights into the factors determining the degree to which a fibre arrangement is suitable for damage to initiate and propagate will help enhance the damage properties of RVEs to withstand higher loads. This could be achieved by implementing slight modifications in the positions of fibres in a given, such that they do not have regions that favour damage initiation and propagation as much as in the initially given RVE.}
\item{Knowledge of factors that decides the degree to which a given arrangement of fibres favours damage in the associated region will assist in identifying the region where damage is anticipated to initiate under a given loading.}
\item{The order in which the individual samples of a given set of RVEs, of a given fibre volume fraction and size, will experience damage initiation can be predicted by evaluating the degree to which the fibre arrangement in each of them favours damage.}
\end{enumerate}
\par Initial investigations in this section are performed on simple RVEs with an almost regular arrangement of fibres by varying just one of the factors anticipated to affect damage initiation. Once the effect of these factors on damage initiation and propagation is identified based on the findings from such analyses, the investigations are done on RVEs with a regular arrangement of fibres by varying these factors together. Once their individual and combined effects on damage initiation and propagation are identified, we utilise the findings to predict damage initiation location in more realistic RVEs with a random arrangement of fibres.
\FloatBarrier
\subsection{Factors Favouring Damage Initiation in 2D Composite RVEs}
\label{sn: Factors Favouring Damage Initiation in 2D Composite RVEs}
Up to this point, more than 1200 RVEs have been considered, and more than 3600 simulations have been performed in this study. Inspecting the damage initiation location, crack propagation, and the crack path in these RVE samples, two significant factors influencing damage initiation in 2D composite RVEs have been identified; 1). minimum \textit{freepath} in RVE, and 2). the angle between the direction of loading and orientation of fibres. 
\FloatBarrier
\subsubsection{Minimum \textit{Freepath} in RVE}
\label{sn: Minimum Freepath in RVE}
\begin{figure}[H]
	\centering
	\includegraphics[width=7.0cm]{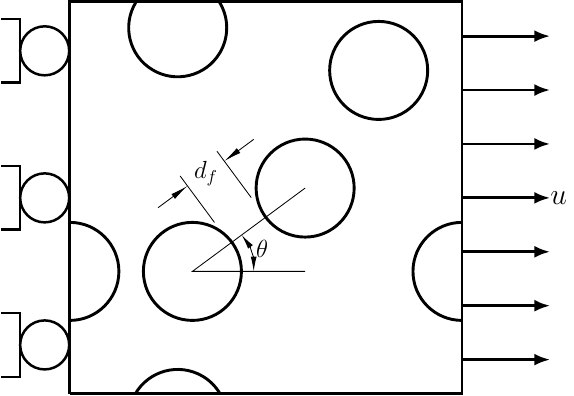}
	\caption{Diagram illustrating freepath $d_{f}$, and angle between a straight line drawn between two fibre centres and the direction of loading, $\theta$. Freepath is shown to be the shortest distance between the boundaries of two fibres, so there is no other fibre boundary in that path.\label{freepath_diagram}}
\end{figure}

Let the minimum distance between the boundaries of two fibres in an RVE, without a third fibre boundary in between them, be termed as \textit{freepath} (\figurename~\ref{freepath_diagram}). It has been observed that damage initiates at regions where this freepath is minimum. It can be attributed to the fact that stress concentration will be higher between closely placed fibres, leading to crack formation. 
\par To study the influence of minimum freepath on damage initiation, an RVE of size 1 unit $\times$ 1 unit. Two studies are carried out using two samples: (1) 100 fibres in the RVE, arranged into 10 rows and 10 columns, and the radius of each fibre $r$ is 0.03 units equivalent to a fibre volume fraction of 0.30, and (2) with 625 fibres occupying a total volume fraction of 0.25 is considered\footnote{Since the fibres are defined by selecting elements, rather than by defining geometric boundaries, the fibres in this particular set of RVE samples are square in shape. The mesh size, 0.005 units, is significant in this context, considering the intended radius of the fibres, 0.01 units, leading to the square shapes of fibres. Radius $r$ in these samples would mean half the width of the fibre.} These fibres are arranged into 25 rows and 25 columns. 
\par These RVE samples are modelled in ABAQUS and are discretised into finite elements of size 0.005 units. They are loaded as in the $xx$ loading condition, with a displacement of 0.50 units applied on the right edge such that it varies from 0 to 0.50 units in $2\times10^{5}$ increments. The samples are analysed, prescribing displacement PBCs, and force-displacement data is obtained. 
 \begin{figure}[H]
 	\centering
 	\includegraphics[width = \textwidth]{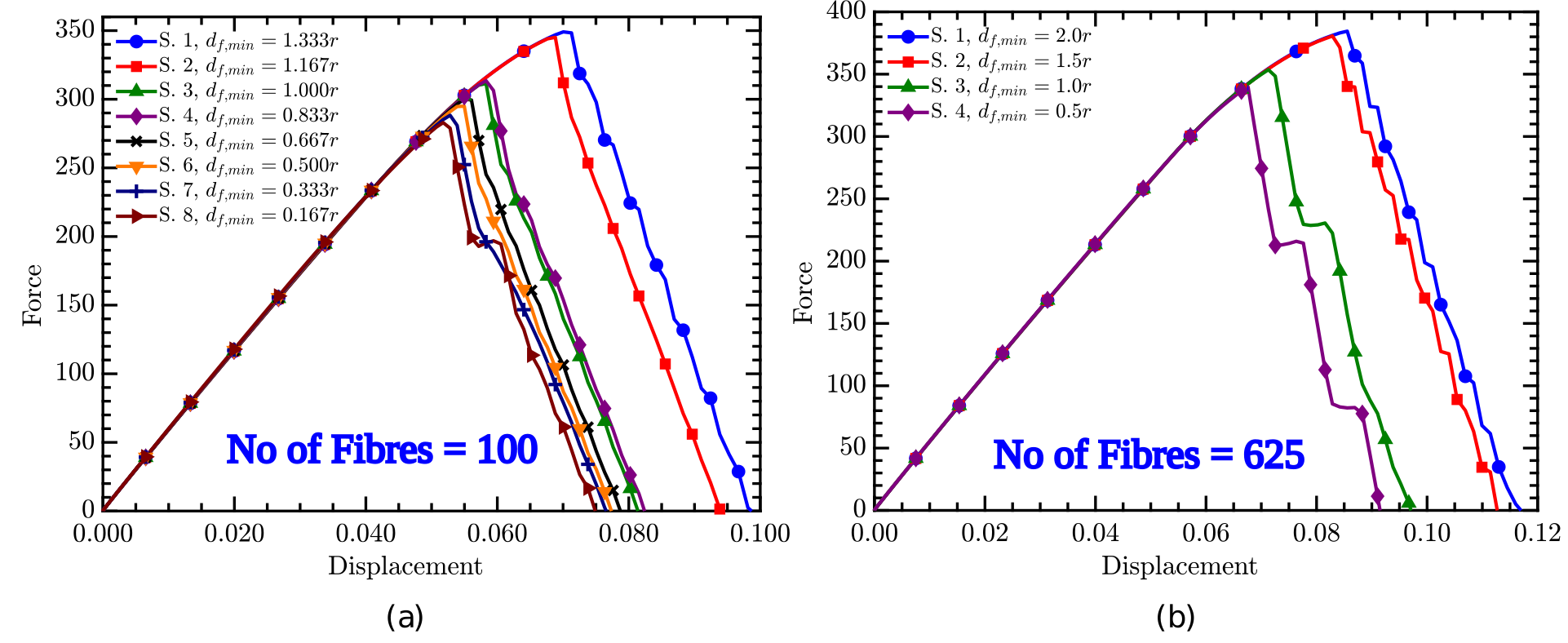}
	\caption{$F-d$ curves for the RVE samples considered to study the influence of minimum freepath on damage initiation strain and failure strain (a) with 100 and (b) 625 inhomogeneities.\label{Dist_100_dist_vs_dmg_ini0}}
 \end{figure}
 \par It has been observed that, in each of these samples, damage initiation always occurs between the pairs of fibres where the freepath is minimum. This suggests that \textbf{a lesser freepath makes the corresponding location more favourable for damage}. The $F$-$d$ curves of these samples are plotted in \figurename~\ref{Dist_100_dist_vs_dmg_ini0}. It can be inferred from \figurename~\ref{Dist_100_dist_vs_dmg_ini0}(a) and \ref{Dist_100_dist_vs_dmg_ini0}(b) that damage initiation occurs at a lesser value of strain with the decrease in minimum freepath in the RVE. Also, with the decrease in minimum freepath, complete failure is expected to occur at a lesser strain. It has been found that a decrease in the minimum freepath by 0.035 units (or by 1.167 times the fibre radius) caused a decrease in damage initiation strain by 26.111\% and a decrease in failure strain by 24.621\% for sample 1 and a decrease in the minimum freepath by 0.015 units (or by 1.5 times the fibre radius) decreased the damage initiation strain and failure strain by 10.782\% and 21.384\% for sample 2, respectively.
 \par
\begin{figure}[H]
	\centering
 	\includegraphics[width = 0.9\textwidth]{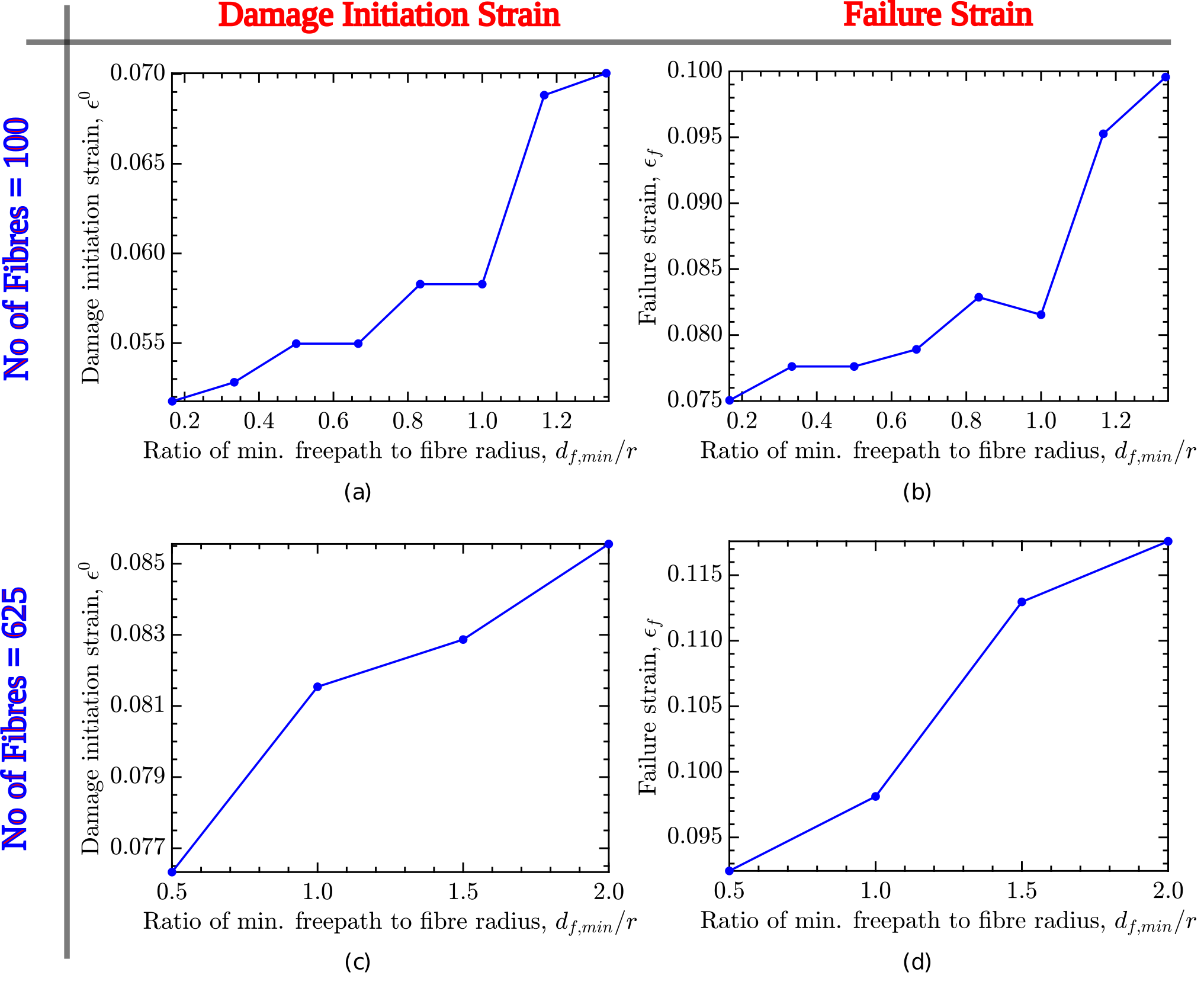}
	\caption{Damage initiation strain shown in (a) and (c) for RVE samples with 100 and 625 inhomogeneities, respectively, plotted against the ratio of minimum freepath to fibre radius. Failure strain is shown in (b) and (d) for RVE samples with 100 and 625 inhomogeneities plotted against the ratio of minimum freepath to fibre radius. \label{Dist_100_dist_vs_dmg_ini}}
\end{figure}

 The two key findings from this study are as follows:
 \begin{enumerate}
 	\item{Closer placement of fibres makes the region more favourable for damage, and the existence of a better such potential region prone to damage lowers the damage initiation strain of the RVE (\figurename~\ref{Dist_100_dist_vs_dmg_ini}).}
 	\item{A decrease in minimum freepath in the RVE lowers the failure strain as well (\figurename~\ref{Dist_100_dist_vs_dmg_ini}(a) and \ref{Dist_100_dist_vs_dmg_ini}(b)).}
 \end{enumerate}
\subsubsection{Angle Between Direction of Loading and Orientation of Fibres}
\label{sn: Angle Between Direction of Loading and Orientation of Fibres} 
\begin{figure}[H]
 	\centering
 	\includegraphics[width = 7.0cm]{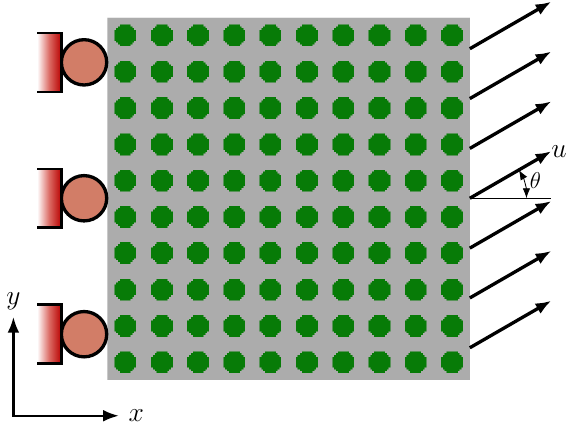}
	\caption[An RVE sample with 100 inhomogeneities with a displacement $u$ prescribed on the right edge, at an angle $\theta$.]{An RVE sample with 100 inhomogeneities with a displacement $u$ prescribed on the right edge, at an angle $\theta$\label{Angle_loading_RVE_100}}
 \end{figure}

The angle between the direction of loading and a line joining the centre of two fibres (\figurename~\ref{freepath_diagram}) is seen to have a prominent effect on damage initiation. Sample 1 of the set of RVEs with 100 inhomogeneities and sample 2 with 625 inhomogeneities are considered to verify the same. The same is loaded as shown in \figurename~\ref{Angle_loading_RVE_100}. 
\par The RVE is constrained on the left edge in the $x$ direction. On the opposite edge, a displacement of 0.5 units\footnote{This displacement is applied by resolving the same into its components. Hence, on each node on the right edge, a displacement of magnitude 0.5 cos $\theta$ and 0.5 sin $\theta$ are applied in the $x$ and $y$ directions, respectively.} is applied at an angle $\theta$ concerning the positive $x$ axis. The displacement is applied such that it varies from 0 to 0.50 units in $2\times10^{5}$ increments. In this combination of the RVE sample and loading condition, the angle between the fibres' orientation and the loading direction is $\theta$, whose influence is to be investigated further.
\par In the first analysis, $\theta$ is kept as 0$^\circ$. In further analyses, $\theta$ is changed to 5$^\circ$, 10$^\circ$, 15$^\circ$, and so on up to 80$^\circ$. Hence, a total of 17 simulations are to be performed for both samples.
\par
\begin{figure}[H]
	\centering
 	\includegraphics[width = 1\textwidth]{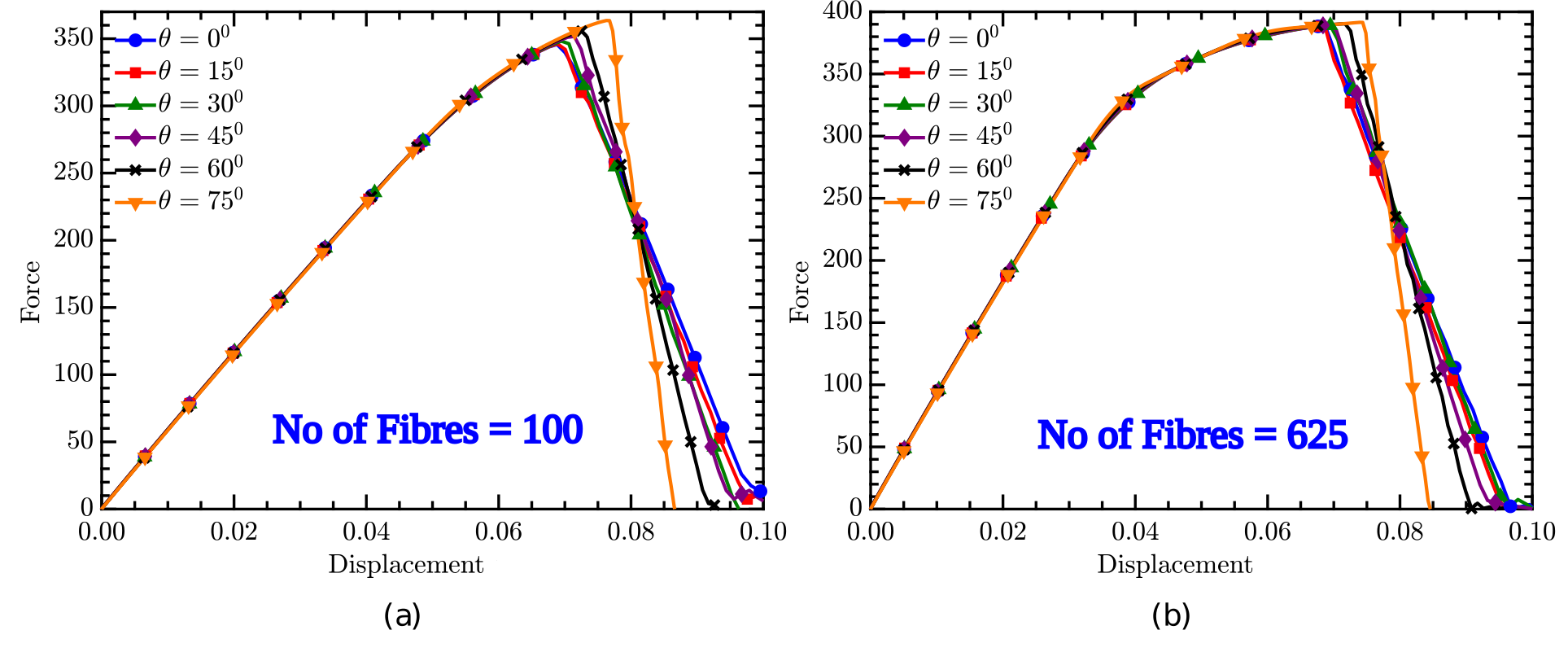}
	\caption{$F-d$ curves for the RVE samples considered to study the influence of angle, $\theta$ (angle between the direction of loading and orientation of fibres) on damage initiation strain and failure strain (a) with 100 and (b) 625 inhomogeneities.}
	\label{fd_angle_100}
\end{figure}

 The $F$-$d$ curves of some selected analyses are plotted in \figurename~\ref{fd_angle_100}. As the angle between the direction of loading and orientation of fibres ($\theta$) increases, damage initiation strain increases, whereas failure strain decreases. It is observed that for sample 1, an increase in $\theta$ by 80$^\circ$\ has increased the damage initiation strain by 14.640\%, and decreased the failure strain by 19.731\% and for sample 2, an increase in $\theta$ by 80$^\circ$\ has increased the damage initiation strain by 14.019\% and decreased the failure strain by 16.349\%.
\par Similarly, the same set of analyses is performed on sample 1 of the set of RVEs with 625 inhomogeneities and force-displacement data are obtained. \figurename~\ref{fd_angle_100}(b) presents the $F$-$d$ curves of some selected analyses. The observations from the $F$-$d$ curves in the previous analyses are also valid.  
\par
\begin{figure}[H]
	\centering
 	\includegraphics[width = 1\textwidth]{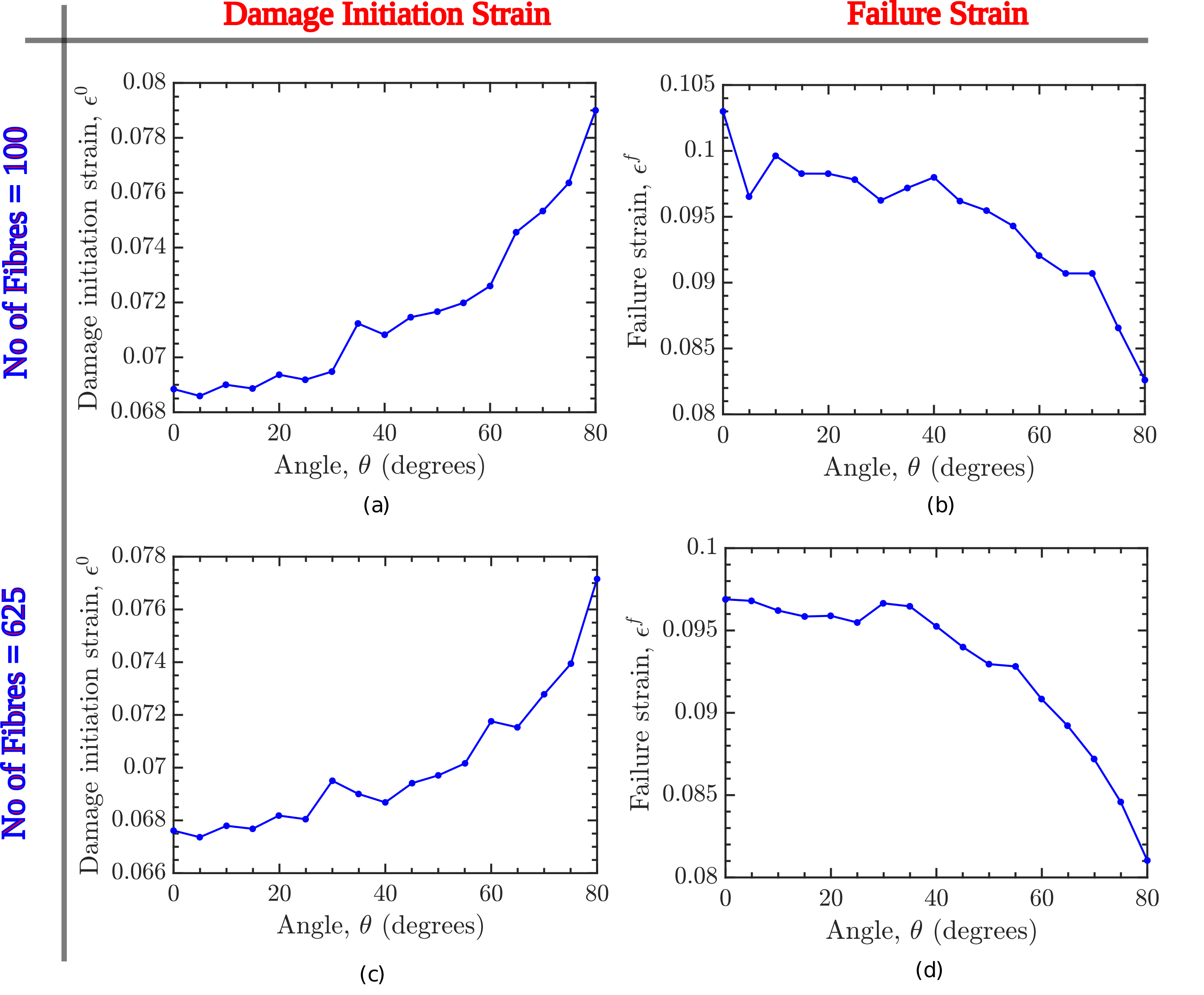}
	\caption{Damage initiation strain shown in (a) and (c) for RVE samples with 100 and 625 inhomogeneities, respectively, plotted against angle, $\theta$ (angle between the direction of loading and orientation of fibres). Failure strain shown in (b) and (d) for RVE samples with 100 and 625 inhomogeneities, respectively, plotted against angle, $\theta$ (angle between the direction of loading and line joining the centres of fibres).}
	\label{Dist_625_angle_vs_dmg_ini}
\end{figure}

 \figurename~\ref{Dist_625_angle_vs_dmg_ini}\hyperref[Dist_625_angle_vs_dmg_ini]{(a)} shows the damage initiation strains in each of these analyses performed on the two RVE samples (one with 100 inhomogeneities, and another with 625 inhomogeneities) plotted against $\theta$ in the corresponding analysis. It establishes that an increase in the angle between the loading direction and fibres' orientation raises the strain at which damage onsets. Hence, \textbf{when two fibres are positioned adjacent in an RVE such that the angle between a line joining their centres and the direction of loading is small, then the location between these fibres potentially favours damage to occur there}.
\par Failure strain in each of these analyses on RVE samples with 100, as well as 625 inhomogeneities, are plotted against $\theta$ in the corresponding analysis, in \figurename~\ref{Dist_625_angle_vs_dmg_ini}\hyperref[Dist_625_angle_vs_dmg_ini]{(b)}. The change in $\theta$ does not noticeably show its effect on failure strain from 0$^\circ$\ to 40$^\circ$. But, from around 40$^\circ$, there is a comparatively steep decline in failure strain.
\par \tablename~\ref{inc_dec} summarises the key findings from this section until this point. The takeaway from this section is that a location inside the RVE would favour damage more if two fibres are positioned there such that they are closer to each other and/or the angle between the direction of loading and their orientation is small.
\begin{table}[H]
	\centering
	\caption{Variation in damage initiation strain, and failure strain, with the variation in minimum freepath $d_{f,min}$, and the angle between the direction of loading and the direction of orientation of fibres $\theta$.\label{inc_dec}}
	\begin{tabular}{@{}lll@{}}
		\toprule
		\begin{tabular}[c]{@{}l@{}}\end{tabular} & \begin{tabular}[c]{@{}l@{}}Damage initiation\\ strain, $\varepsilon^{0}$\end{tabular} & \begin{tabular}[c]{@{}l@{}}Failure strain,\\ $\varepsilon_{f}$\end{tabular} \\ \midrule
		$d_{f,min}$ increases $(\uparrow)$ & increases ($\uparrow$)  & increases ($\uparrow$)\\
		$\theta$ increases ($\uparrow$) & increases ($\uparrow$)  & decreases ($\downarrow$)\\ \bottomrule
	\end{tabular}
\end{table} 

\subsection{Relative Contributions of Change in Minimum Freepath and Change in Angle Between Direction of Loading and Orientation of Fibres, to the Change in Damage Initiation Strain}
\label{Sn: Relative Contributions of Change in Minimum Freepath and Change in Angle Between Direction of Loading and Orientation of Fibres, to the Change in Damage Initiation Strain}
As seen in sections \ref{sn: Minimum Freepath in RVE} and \ref{sn: Angle Between Direction of Loading and Orientation of Fibres}, a change in minimum freepath and angle between the direction of loading and direction of fibre orientation will induce a change in damage initiation strain. This section aims to quantify the relative contributions of change in these factors to the consequent change in damage initiation strain. In other words, this section investigates what percentage of the change in damage initiation strain is due to a change in $d_{f,min}$ and what percentage of the same is due to a change in $\theta$.
\par The eight RVE samples (with 100 inhomogeneities) considered to study the influence of freepath on damage initiation strain are utilised in this investigation. So, we have eight RVE samples in this study with $d_{f,min}$ ranging from 0.04 units (or 1.1333$r$) to 0.005 units (or 0.167$r$). Each of these samples is loaded as in \figurename~\ref{Angle_loading_RVE_100}. Each of these samples is analysed by prescribing the value of $\theta$ as 0$^\circ$, 10$^\circ$, 20$^\circ$, 30$^\circ$, 40$^\circ$, 50$^\circ$, 60$^\circ$, 70$^\circ$, and 80$^\circ$. That is, for each of these samples, nine analyses are performed. The load, boundary conditions, damage initiation criteria, and damage evolution law are the same as followed in this chapter until now. After the analyses, Force-displacement data is obtained. The damage initiation strains from these analyses are tabulated in \tablename~\ref{dmg_ini_strain_table}.
\par Now that the values of damage initiation strains at different combinations of $d_{f,min}$ and $\theta$ are obtained, we proceed to quantify the relative contributions of the factors. The damage initiation strain from the analysis of sample 1 (which has a regular arrangement of fibres), when prescribed $\theta$ as 0$^\circ$ is kept as the reference. In the rest of the analyses, $d_{f,min}$ and/or $\theta$ change with respect to those in sample 1. Also, the damage initiation strains change. The new value of damage initiation strain is considered a consequence of changes in $d_{f,min}$ and $\theta$. 
\begin{table}[H]
	\caption[Damage initiation strain ($\times10^{-2}$) at each combination of $d_{f,min}$ and $\theta$.]{Damage initiation strain ($\times10^{-2}$) at each combination of $d_{f,min}$ and $\theta$. The ratio of each of the damage initiation strain values to that of sample 1 (the one among the samples with regular arrangement of fibres, $d_{fmin}$ = 0.04 units or 1.333$r$) when $\theta$ prescribed as 0$^\circ$\ are given in brackets.\label{dmg_ini_strain_table}}
	\centering
		\begin{tabular}{@{}cccccccccc@{}}
			\toprule
			\multicolumn{2}{c}{\multirow{2}{*}{}} & \multicolumn{8}{c}{Minimum freepath $d_{f,min}$}                                                                                                                                                                                                                                                                                                                                                                                                                                                                 \\ \cmidrule(l){3-10} 
			\multicolumn{2}{c}{}& 1.333$r$ & 1.167$r$ & 1.000$r$ & 0.833$r$ & 0.667$r$ & 0.500$r$ & 0.333$r$ &0.167$r$\\ \midrule\vspace{0.5em}
			& 0 & \begin{tabular}[c]{@{}c@{}}6.882\\ (1.000)\end{tabular} & \begin{tabular}[c]{@{}c@{}}6.882\\ (1.000)\end{tabular} & \begin{tabular}[c]{@{}c@{}}5.829\\ (0.847)\end{tabular} & \begin{tabular}[c]{@{}c@{}}5.829\\ (0.847)\end{tabular} & \begin{tabular}[c]{@{}c@{}}5.497\\ (0.799)\end{tabular} & \begin{tabular}[c]{@{}c@{}}5.497\\ (0.799)\end{tabular} & \begin{tabular}[c]{@{}c@{}}5.282\\ (0.767)\end{tabular} & \begin{tabular}[c]{@{}c@{}}5.176\\ (0.752)\end{tabular} \\\vspace{0.5em}
			& 10 & \begin{tabular}[c]{@{}c@{}}6.898\\ (1.002)\end{tabular} & \begin{tabular}[c]{@{}c@{}}6.898\\ (1.002)\end{tabular} & \begin{tabular}[c]{@{}c@{}}5.851\\ (0.850)\end{tabular} & \begin{tabular}[c]{@{}c@{}}5.851\\ (0.850)\end{tabular} & \begin{tabular}[c]{@{}c@{}}5.521\\ (0.802)\end{tabular} & \begin{tabular}[c]{@{}c@{}}5.414\\ (0.787)\end{tabular} & \begin{tabular}[c]{@{}c@{}}5.307\\ (0.771)\end{tabular} & \begin{tabular}[c]{@{}c@{}}5.202\\ (0.756)\end{tabular} \\\vspace{0.5em}
			\multirow{9}{*}{\rotatebox[origin=c]{90}{$\theta$ (degrees)}}& 20     & \begin{tabular}[c]{@{}c@{}}6.934\\ (1.007)\end{tabular} & \begin{tabular}[c]{@{}c@{}}6.816\\ (0.990)\end{tabular} & \begin{tabular}[c]{@{}c@{}}5.798\\ (0.842)\end{tabular} & \begin{tabular}[c]{@{}c@{}}5.907\\ (0.858)\end{tabular} & \begin{tabular}[c]{@{}c@{}}5.583\\ (0.811)\end{tabular} & \begin{tabular}[c]{@{}c@{}}5.477\\ (0.796)\end{tabular} & \begin{tabular}[c]{@{}c@{}}5.372\\ (0.781)\end{tabular} & \begin{tabular}[c]{@{}c@{}}5.269\\ (0.766)\end{tabular} \\\vspace{0.5em}
			& 30 & \begin{tabular}[c]{@{}c@{}}6.947\\ (1.009)\end{tabular} & \begin{tabular}[c]{@{}c@{}}6.947\\ (1.009)\end{tabular} & \begin{tabular}[c]{@{}c@{}}5.855\\ (0.851)\end{tabular} & \begin{tabular}[c]{@{}c@{}}5.960\\ (0.866)\end{tabular} & \begin{tabular}[c]{@{}c@{}}5.648\\ (0.821)\end{tabular} & \begin{tabular}[c]{@{}c@{}}5.545\\ (0.806)\end{tabular} & \begin{tabular}[c]{@{}c@{}}5.344\\ (0.776)\end{tabular} & \begin{tabular}[c]{@{}c@{}}5.244\\ (0.762)\end{tabular} \\\vspace{0.5em}
			& 40 & \begin{tabular}[c]{@{}c@{}}7.081\\ (1.029)\end{tabular} & \begin{tabular}[c]{@{}c@{}}6.974\\ (1.013)\end{tabular} & \begin{tabular}[c]{@{}c@{}}5.946\\ (0.864)\end{tabular} & \begin{tabular}[c]{@{}c@{}}5.946\\ (0.864)\end{tabular} & \begin{tabular}[c]{@{}c@{}}5.652\\ (0.821)\end{tabular} & \begin{tabular}[c]{@{}c@{}}5.556\\ (0.807)\end{tabular} & \begin{tabular}[c]{@{}c@{}}5.366\\ (0.780)\end{tabular} & \begin{tabular}[c]{@{}c@{}}5.272\\ (0.766)\end{tabular} \\\vspace{0.5em}
			& 50 & \begin{tabular}[c]{@{}c@{}}7.163\\ (1.041)\end{tabular} & \begin{tabular}[c]{@{}c@{}}6.969\\ (1.013)\end{tabular} & \begin{tabular}[c]{@{}c@{}}5.942\\ (0.863)\end{tabular} & \begin{tabular}[c]{@{}c@{}}6.032\\ (0.876)\end{tabular} & \begin{tabular}[c]{@{}c@{}}5.674\\ (0.824)\end{tabular} & \begin{tabular}[c]{@{}c@{}}5.587\\ (0.812)\end{tabular} & \begin{tabular}[c]{@{}c@{}}5.413\\ (0.786)\end{tabular} & \begin{tabular}[c]{@{}c@{}}5.327\\ (0.774)\end{tabular} \\\vspace{0.5em}
			& 60 & \begin{tabular}[c]{@{}c@{}}7.254\\ (1.054)\end{tabular} & \begin{tabular}[c]{@{}c@{}}7.087\\ (1.030)\end{tabular} & \begin{tabular}[c]{@{}c@{}}6.035\\ (0.877)\end{tabular} & \begin{tabular}[c]{@{}c@{}}6.114\\ (0.888)\end{tabular} & \begin{tabular}[c]{@{}c@{}}5.802\\ (0.843)\end{tabular} & \begin{tabular}[c]{@{}c@{}}5.725\\ (0.832)\end{tabular} & \begin{tabular}[c]{@{}c@{}}5.496\\ (0.799)\end{tabular} & \begin{tabular}[c]{@{}c@{}}5.421\\ (0.788)\end{tabular} \\\vspace{0.5em}
			& 70 & \begin{tabular}[c]{@{}c@{}}7.527\\ (1.094)\end{tabular} & \begin{tabular}[c]{@{}c@{}}7.147\\ (1.038)\end{tabular} & \begin{tabular}[c]{@{}c@{}}6.151\\ (0.894)\end{tabular} & \begin{tabular}[c]{@{}c@{}}6.212\\ (0.903)\end{tabular} & \begin{tabular}[c]{@{}c@{}}5.907\\ (0.858)\end{tabular} & \begin{tabular}[c]{@{}c@{}}5.787\\ (0.841)\end{tabular} & \begin{tabular}[c]{@{}c@{}}5.607\\ (0.815)\end{tabular} & \begin{tabular}[c]{@{}c@{}}5.547\\ (0.806)\end{tabular} \\\vspace{0.5em}
			& 80 & \begin{tabular}[c]{@{}c@{}}7.907\\ (1.149)\end{tabular} & \begin{tabular}[c]{@{}c@{}}7.401\\ (1.075)\end{tabular} & \begin{tabular}[c]{@{}c@{}}6.364\\ (0.925)\end{tabular} & \begin{tabular}[c]{@{}c@{}}6.448\\ (0.937)\end{tabular} & \begin{tabular}[c]{@{}c@{}}6.104\\ (0.887)\end{tabular} & \begin{tabular}[c]{@{}c@{}}6.016\\ (0.874)\end{tabular} & \begin{tabular}[c]{@{}c@{}}5.866\\ (0.852)\end{tabular} & \begin{tabular}[c]{@{}c@{}}5.744\\ (0.835)\end{tabular} \\ \bottomrule
		\end{tabular}
\end{table}

\begin{table}[H]
	\caption[Damage initiation strain ($\times10^{-2}$) at each combination of $d_{f,min}$ and $\theta$.]{Percentage contribution of $\theta$ to change in damage initiation strain $c_{\theta}$ at various combinations of variation of $(d_{f,min})_{ref}$ and $\theta$. Each row corresponds to the increase in $\theta$ from that of the reference analysis, $\theta_{ref}$, which is 0$^\circ$. Each column corresponds to the decrease in $d_{f,min}$ from that of the reference analysis, $(d_{f,min})_{ref}$, which is 1.333$r$ or 0.04 units. Each cell in the table tells us the percentage contribution of change in angle to change in damage initiation due to a combined change of minimum freepath by $(d_{f,min})^{\prime}-(d_{f,min})_{ref}$ and angle between the direction of loading and direction of fibre orientation by $\theta_{ref} - \theta^{\prime}$.\label{dmg_ini_c_theta_table}}
	\centering
		\begin{tabular}{@{}cccccccccc@{}} 
			\toprule
			\multicolumn{2}{c}{\multirow{2}{*}{}} & \multicolumn{8}{c}{$\left((d_{f,min})_{ref}-(d_{f,min})^{\prime}\right)/r$}\\ \cmidrule(l){3-10} 
			\multicolumn{2}{c}{}& 0 & 0.167 & 0.333 & 0.500 & 0.667 & 0.833 & 1.00 &1.167\\ \midrule
			\multirow{9}{*}{\rotatebox[origin=c]{90}{$\theta_{ref} - \theta^{\prime}$ (degrees)}} 
			& 0 &  & 0.000    & 0.000  & 0.000 & 0.000  & 0.000  & 0.000  & 0.000  \\
			& 10 & 100.000 & 100.000  & 2.098  & 2.098  & 6.838  & -5.961 & 1.563  & 1.497  \\
			& 20 & 100.000 & -130.611 & -2.777 & 7.089 & 5.979 & -1.395 & 5.480  & 5.275  \\
			& 30 & 100.000 & 100.000  & 2.365  & 11.763 & 10.376 & 3.319 & 3.704 & 3.850  \\
			& 40 & 100.000 & 46.231 & 9.344  & 9.344  & 9.788 & 3.714 & 4.679 & 5.060  \\
			& 50 & 100.000 & 30.850 & 8.486  & 15.260 & 10.639 & 5.363 & 6.954 & 7.591  \\
			& 60 & 100.000 & 55.065 & 14.490 & 20.006 & 17.332 & 12.949 & 10.872 & 11.795 \\
			& 70 & 100.000 & 41.023 & 18.967 & 22.574 & 20.202 & 14.252 & 14.470 & 15.792 \\
			& 80 & 100.000 & 50.627 & 25.760 & 29.818 & 25.197 & 21.520 & 22.255 & 20.822 \\ \bottomrule 
		\end{tabular}
\end{table}
\par Let us consider a case when $d_{f,min}$ is 0.833$r$ and $\theta$ is 30$^\circ$. As mentioned earlier, the reference is sample 1 ($d_{f,min} = 1.333r$) analysed at $\theta$ = 0$^\circ$. When $d_{f,min}$ alone changes from 1.333$r$ to 0.833$r$, damage initiation strain changes from $6.822\times10^{-2}$ to $5.829\times10^{-2}$. When $\theta$ alone changes from 0$^\circ$ to 30$^\circ$, damage initiation strain changes from $6.822\times10^{-2}$ to $6.947\times10^{-2}$. The damage initiation strain when $d_{f,min}$ is 0.833$r$ and $\theta$ is 30$^\circ$\ is considered to be
\begin{equation}
	\varepsilon^{0}_{0.833r,\:30^{\circ}} = \left(\dfrac{c_{\theta}}{100}\right)\varepsilon^{0}_{1.333r,\:30^{\circ}} + \left(1-\dfrac{c_{\theta}}{100}\right)\varepsilon^{0}_{0.833r,\:0^{\circ}}
	\label{eqn_angle_contribution}
\end{equation}
where $c_{\theta}$ is the percentage contribution of the change in $\theta$ to the change in damage initiation strain. All the terms in Eq. \ref{eqn_angle_contribution} are known except $c_{\theta}$. Hence by solving this equation, $c_{\theta}$ can be found for the particular case. That is, we can detect what percentage of change in damage initiation strain is due to change in angle when $d_{f,min}$ is changed from 1.333$r$ to 0.833$r$, and $\theta$ is changed from 0$^\circ$\ to 30$^\circ$. Once $c_{\theta}$ is obtained, the percentage contribution of change in $d_{f,min}$ to change in damage initiation strain could be found out as (100 - $c_{\theta}$). Using the framework for quantification of individual contributions of change in $d_{f,min}$ and $\theta$ to the change in damage initiation strain from that of the reference analysis, the percentage contributions of $d_{f,min}$ and $\theta$ to change in damage initiation strain at various analyses performed in this investigation are found. They are presented in \tablename~\ref{dmg_ini_c_theta_table}.
\begin{figure}[H]
	\centering
	\includegraphics[trim = 0 0 0 0, clip,width=0.75\textwidth]{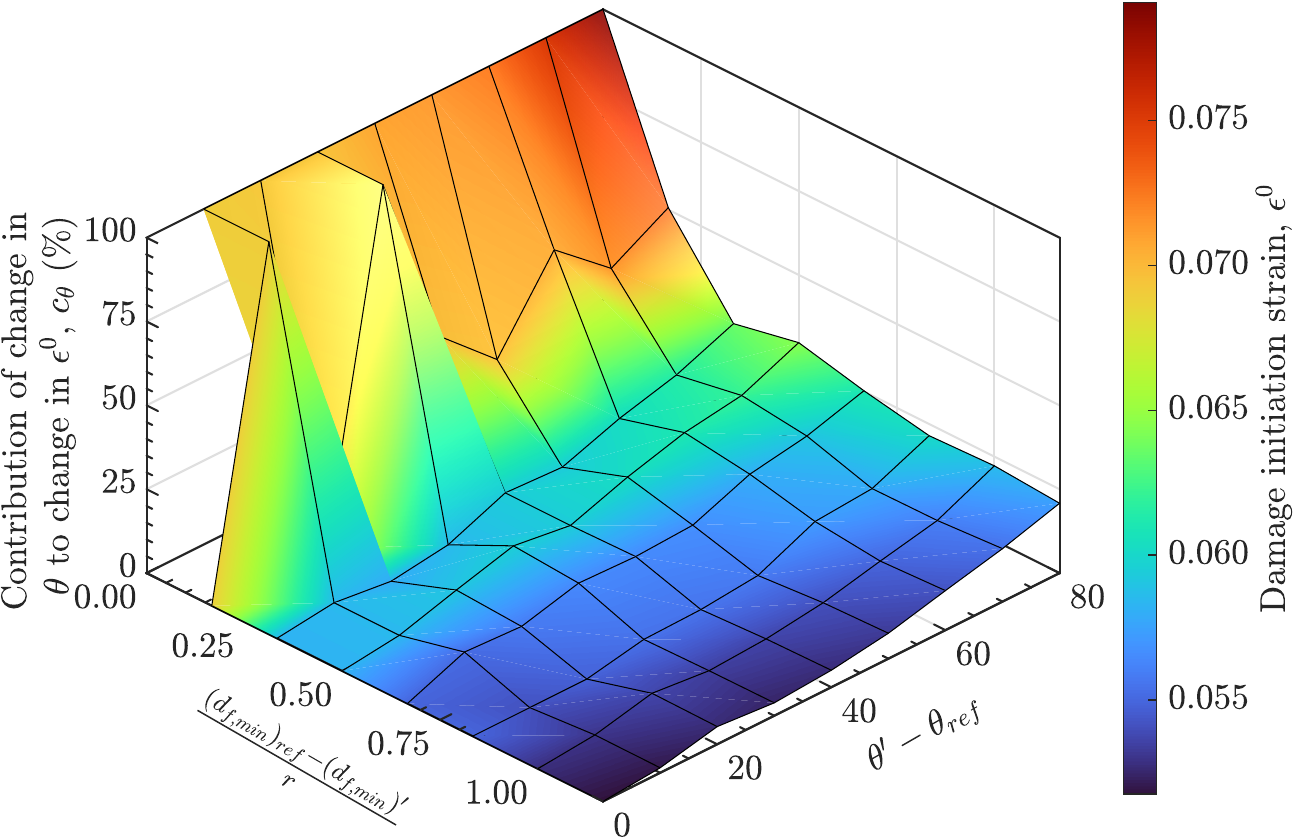}
	\caption[Surface plot showing the contribution of change in angle to variation in damage initiation strain.]{Surface plot showing the contribution of change in angle to variation in damage initiation strain. This plot summarises the important findings in this section on factors favouring damage initiation in 2D composite RVEs. It can be observed that once the change in $d_{f,min}$ goes above a small limit, the relative contribution of change in $\theta$ decreases, and that of change in $d_{f,min}$ becomes considerably high. The colour at each data point corresponds to the damage initiation strain corresponding to that combination of $(d_{f,min})^{\prime}$ and $\theta^{\prime}$. As we move along the $x$ axis from its zero, the value of $d_{f,min}^{\prime}$ decreases. It can be observed that as the minimum freepath decreases, the damage initiation strain also decreases. As we move along the $y$ axis from its zero, the value of $\theta$ increases, and so does the damage initiation strain. Also, as $(d_{f,min})$ and $\theta$ increase, damage initiation strain increases. \label{surface_plot_ca}}
\end{figure}
\par The surface plot in \figurename~\ref{surface_plot_ca} presents the functional relationship between the dependent variable $c_{\theta}$, plotted on the $z$ axis, and the two independent variables—variations in $d_{f,min}$ and $\theta$ from that of the reference analysis, are plotted on the $x$ and $y$ axis, respectively. It can be inferred that once the change in minimum freepath goes above 0.167$r$, the effect of change in angle on change in damage initiation strain from that corresponding to the reference analysis diminishes to the range 0-30\%. This means that once $d_{f,min}$ starts to vary by a noticeable value, its effect will dominate over the same due to change in $\theta$. Therefore, variation in minimum freepath is a more determining factor than the angle variation between the loading direction and fibres' orientation in damage initiation in composite RVEs.
\section{Conclusions}
\label{Conclusions}
This research aimed to increase the representativeness of an RVE by attenuating the mesh and size sensitivities in their modelling and identifying the factors favouring damage initiation in 2D composites. It can be concluded that the variation in $F$-$d$ curves of a material subjected to a given loading analysed at different mesh sizes is due to the localisation of strain to a region of width equal to the mesh size. A technique is proposed to alleviate mesh sensitivity in RVE modelling. It has been demonstrated that an RVE when analysed under a given load but with different mesh sizes, would yield the same $F$-$d$ curve if the product of failure strain and the square root of mesh size in all the analyses are the same. The proposed technique to alleviate mesh sensitivity aids in obtaining the material response independent of the bandwidth of strain localisation. It also assists in performing analyses utilising a coarser mesh and still results in material response, as observed from analysis with a finer mesh.
\par After analysing the $F$-$d$ curves for same-size RVE samples of several fibre volume fractions, it has been observed that the post-damage dispersion of $F$-$d$ curves is lower at higher volume fractions. This tells us that the effects of localisation of damage due to the position of heterogeneities on material response decrease with an increase in fibre volume fraction and that the results from different RVEs of the same size become more precise at higher fibre volume fractions. Further, upon repeating the experiment at other RVE sizes, it has been found that this trend of decrease in post-damage dispersion of $F$-$d$ curves with the increase in fibre volume fraction persists irrespective of the RVE size. 
\par Comparing the post-damage dispersion of $F$-$d$ curves of RVE samples of the same fibre volume fraction but of different RVE sizes, two important conclusions are drawn: material response from larger RVEs will be more precise for a given fibre volume fraction; use of larger RVEs will be beneficial if the fibre volume fraction is less than or equal to 0.50. For analysing RVEs of fibre volume fraction above that, a relatively smaller RVE would suffice to get precise results.
\par  This study clearly shows that the effect of RVE size on material response is lesser at higher fibre volume fractions. A set of modified PBCs (MPBC) has been proposed to attenuate the effects of RVE size on the observed material response. MPBCs, in addition to directing the displacements periodically on the opposite nodes on the domain boundaries, prescribe the periodicity of strain across the boundaries. These MPBCs are applied by constraining the nodal displacements on and near the boundaries, hence, eliminating the requirement of directly constraining the strain near the domain boundary. The width of the region near the edge where MPBCs are prescribed dictates the efficacy of MPBCs in reducing size sensitivity in RVE modelling. 
\par Another significant contribution of this work is studying factors favouring damage initiation in 2D composite RVEs. The study identifies that a region in the RVE becomes more favourable to damage if two fibres are placed very close to each other in that region, with the angle between the direction of loading and an imaginary line joining their centres being less. Throughout this work, the minimum distance between the boundaries of two fibres in an RVE, without a third fibre boundary in between them, is referred to as minimum freepath $d_{f,min}$, and the angle between the direction of loading and the direction of orientation of a pair of fibres is referred to as $\theta$. This work presents strong evidence that an increase in either $d_{f,min}$ or $\theta$ will increase the damage initiation strain of RVEs with a regular arrangement of fibres. In such RVEs, an increase in both  $d_{f,min}$ and $\theta$ will increase the damage initiation strain. Also, in such RVEs, the effect of an increase in either of the two factors counters the effect due to a decrease in the other. The effect of an increase in $d_{f,min}$ contributes more to the change in damage initiation strain than an increase in $\theta$.
\par This study demonstrates with examples that it is possible to predict damage initiation location even in RVEs with irregular arrangement of fibres, based on our findings on factors that make a region in the RVE more prone to damage. It is a combination of the freepath between two fibres and the orientation of those two fibres concerning the direction of loading that decides the probability of damage initiation in the region between them. The lesser the values of $d_{f,min}$ and $\theta$, the more prone the region is to damage. 
\par This research has been designed as a humble attempt to understand the rationales of microscale damage, overcome the challenges in the computational analyses of composite failure, and get better insights into microscale damage. There is a definite and profuse horizon to extend this endeavour. The scope of such an extensive study may establish a universal nature of our findings regarding the loading conditions, the shape and type of reinforcements, and the scale of analysis. The conclusions of this study are within the regimes of material behaviour under transverse loading. Further investigations addressing the research questions we proposed, probing how our answers would have been if the loading was in the fibre direction, would add to the breadth of this research. We have considered the damage in the constitutive phases of the composite but not in the interfaces of the constitutive phases. In this regard, a study evaluating the effects of damage in the fibre-matrix interface will also add to the robustness of our findings. This study has been performed on composite RVEs with circular fibre reinforcements. Insights into the effect of the shape of the fibres will assist in overcoming challenges in the computational simulation and understanding the damage in further applications. Similarly, performing the microscale damage analysis of composites with other types of reinforcements, such as particulate reinforcements, will be absorbing and meaningful.

\section*{Acknowledgement}
This work was supported by a research grant (ARDB/01/1051983/M/I, Project No. 1983) of the Aeronautical Research \& Development Board, India. The content is solely the responsibility of the authors and does not necessarily represent the official views of the funding organisation.
\color{black}
\section*{Availability of data and materials} The data and materials supporting this study's findings are available from the corresponding author upon reasonable request. 
\appendix
\section{RVE Generation with Random Orientation of Fibres}
The first step of RVE modeling is to define the RVE geometry. Random Sequential Adoption (RSA) technique - a technique originally proposed by \cite{widom-1966}, is the most widely used method for generating microstructure with randomly placed heterogeneities \citep{zhou2016generation, pan2008analysis, bailakanavar2014automated, kari2007computational, kari2007numerical, bohm2002multi, pan2008numerical} and same is adopted in this work. The tool used in our work for obtaining the fibre centre coordinates using RSA technique is MATLAB$^{\circledR}$. The objective addressed in the MATLAB code is to create a specific number of non-intersecting circles inside a square of a given dimension (length = 1 unit in this specific case). The radius of the fibres is decided upon the fibre volume fractions to be achieved and the number of fibres to be placed inside the RVE. 
\par A concept associated with the periodicity of material basically refers to the continuity of the inclusions at the borders of the RVEs. The size of the RVEs to predict the material response with a given level of accuracy is much larger without incorporating this periodicity in terms of inclusion-continuity \citep{Gitman-2007}. The absence of material continuity across the RVE boundaries is also named as \textit{"wall-effect"}. 
\par
\begin{figure}[H]
\centering
	\includegraphics[width=0.375\textwidth]{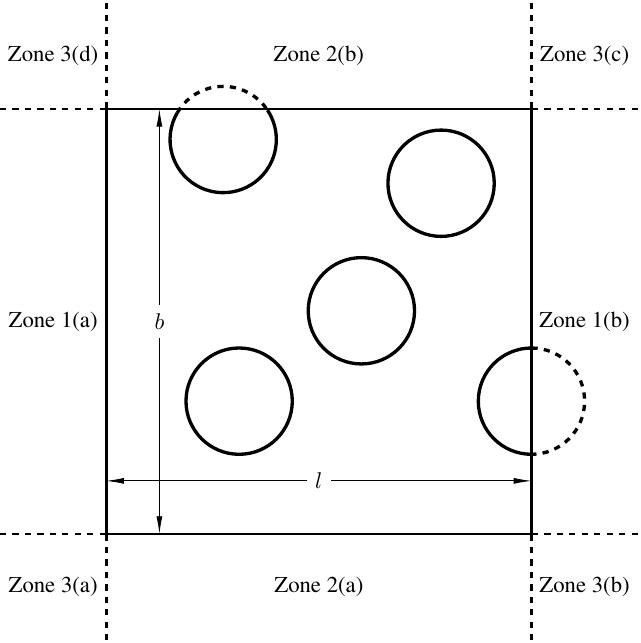}
\caption[Diagram to illustrate the algorithm to eliminate wall-effects]{Diagram to illustrate the algorithm to eliminate wall-effects. An RVE of length $l$ and breadth $b$ is shown. The solid straight lines indicate the boundary of the RVE, and the dashed straight lines separate the different zones around the RVE to demonstrate the algorithm to implement material periodicity. The RVE boundaries are cutting some of the circular fibres, excluding some portion of these fibres from the RVE. These excluded fibre portions are to be included in the RVE such that if the RVE is repeated in both directions in space, there will not be any trace of fibre discontinuity. This eliminates the discontinuity of the inclusions at the borders of the RVEs.}
\label{zones}
\end{figure}

 In case a fibre of radius $r$ is being cut by the RVE boundary, it could either fall into zone 1, 2, or 3 as shown in \figurename~\ref{zones}. Let the length and breadth of the RVE be $l$ and $b$, respectively. Let the centre coordinates of the fibre be ($x,\ y$). If a portion of the fibre falls into zone 3, the corresponding coordinates of the fibre in zone 3(c) is found out by changing $x$ to $x+l$ if $x < r$, and $y$ to $y+l$ if $y < r$. Let these updated values of $x$ and $y$ be $x^{\prime}$ and $y^{\prime}$, respectively. The centre coordinates of the fibre is changed to the updated ones. Further, three new circles are drawn with center coordinates as $(x^{\prime}-l,\ y^{\prime}-b), (x^{\prime}-l,\ y^{\prime}),$ and $(x^{\prime},\ y^{\prime}-b)$, respectively. Else, if a circle falls into zone 1(a), a new circle is drawn with center coordinates as $(x+l,\ y)$. Else, if a circle falls into zone 2(a), a new circle is drawn with center coordinates as $(x,\ y+b)$. Similarly, circles could be mapped from zone 2(b) to 2(a), and 1(b) to 1(a), and new circles could be drawn. 
\par
\begin{figure}[H]
\centering
	\includegraphics[width=0.625\textwidth]{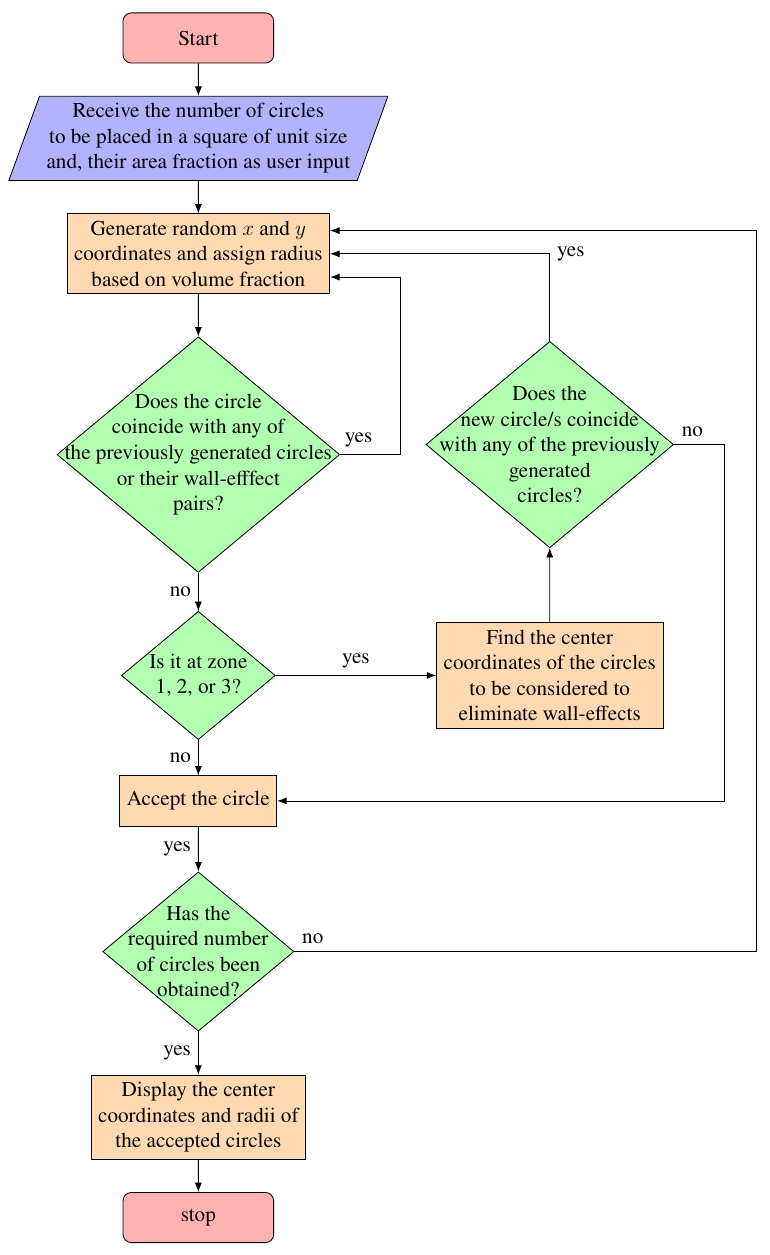}
\caption{Flow diagram explaining the RSA technique for the generation of RVE with randomly orientated fibres}
\label{flowdiagram_rsa}
\end{figure}

 The radius of a newly added fibre should be equal to $r$, which is the radius of that fibre whose continuity across the boundaries is to be ensured by the newly added fibre. After generating a new fibre centre coordinates pair ($x$,\ $y$), the same will be accepted if and only if a circle of the required radius, drawn with its center at ($x$,\ $y$), does not coincide with any of the previously drawn circles and their corresponding \textit{wall-effect} pair/s. The output from the MATLAB code is the center coordinates and the radii of the required number of fibres, and the coordinates of those extra fibres drawn to eliminate wall-effects are not a part of the same. The flow diagram illustrating the algorithm used to generate fibre centre coordinates using the RSA technique while keeping in consideration the \textit{wall-effect} elimination is shown in \figurename~\ref{flowdiagram_rsa}.
\color{black}

\bibliography{bibfile}
\end{document}